\documentclass[a4paper,11pt]{article}
\usepackage{jheppub} 
\usepackage{enumerate}
\usepackage[T1]{fontenc} 
\usepackage{comment,epigraph}
\usepackage{physics}
\usepackage{dsfont,comment}
\usepackage{caption}
\usepackage{slashed}
\usepackage[mathscr]{euscript}
\usepackage{xhfill}

\usepackage[caption=false]{subfig}

\title{Exploring the black hole spectrum of axionic Horndeski theory}

\author[a]{Matteo Baggioli,}
\author[b]{Adolfo Cisterna}
\author[c]{and Konstantinos Pallikaris}

\affiliation[a]{Wilczek Quantum Center, School of Physics and Astronomy, Shanghai Jiao Tong University, Shanghai 200240, China \& Shanghai Research Center for Quantum Sciences, Shanghai 201315.}
\affiliation[b]{Sede Esmeralda, Universidad de Tarapacá,
Av. Luis Emilio Recabarren 2477, Iquique, Chile}
\affiliation[c]{Laboratory of Theoretical Physics, Institute of Physics, University of Tartu, W. Ostwaldi 1, 50411
Tartu, Estonia.}

\emailAdd{b.matteo@sjtu.edu.cn}
\emailAdd{adolfo.cisterna.r@mail.pucv.cl}
\emailAdd{konstantinos.pallikaris@ut.ee}

\abstract{We provide a ``user-friendly'' algorithm to systematically and rapidly obtain exact planar black hole solutions in the Einstein-Maxwell theory deformed by the most general shift- and reflection-symmetric Horndeski sector where the usual Galileon is replaced by a tuple of scalars with profiles linear in the coordinates of the transverse manifold. Under precise assumptions, these axion backgrounds break the translational invariance of the system, causing momentum dissipation in the holographically dual field theory. The success of the method relies on the simple realization that the bulk equations of motion become more tractable when written in terms of the axions kinetic terms, instead of the radial coordinate. Showcasing this particularly efficient recipe, we derive novel asymptotically AdS black holes, and show that their extremal counterparts always flow to an $\mathrm{AdS}_2\times \mathds{R}^2$ infrared fixed point. Additionally, we report an interesting family of new asymptotically Lifshitz black hole solutions with $z>1$. Finally, we discuss the DC transport properties of the dual relativistic field theories in view of possible applications to condensed matter systems.}

\begin{document} 
\maketitle
\flushbottom

\section{Introduction}

With the advent of the Anti-de Sitter/Conformal Field Theory (AdS/CFT) correspondence~\cite{Maldacena:1997re} and the subsequent advances in ``bottom-up'' holography~\cite{CasalderreySolana:2011us,Hartnoll:2009sz}, hairy black holes have become interesting theoretical laboratories offering a fresh insight into the physics of strongly correlated systems, particularly in the context of an AdS/Condensed Matter Theory (AdS/CMT) duality~\cite{Hartnoll:2016apf}. Perhaps one of the first and most prominent examples might as well be the holographic superconductor (or more precisely, superfluid)~\cite{Hartnoll:2008vx} of Hartnoll et al.; a black hole in the bulk, which develops charged hair only at low temperatures, is utilized to successfully describe the formation of a charged condensate in the boundary field theory, and the subsequent superfluid phase transition that takes place. 

Especially relevant when dealing with such gauge/gravity models, \emph{planar black holes}, that is, black holes with a Riemann-flat transverse manifold, have their own place in the ``black'' landscape of gravitational theories. However, precisely because the transverse manifold has vanishing curvature, not all planar solutions describe black holes, and the task of finding those that ultimately do so, is not always a trivial one. In fact, it is a simple exercise in pure General Relativity (GR), to show that any planar solution inevitably describes a naked central singularity, unless a negative cosmological constant is introduced. 

In \cite{Bardoux:2012aw}, a detailed analysis has been performed in order to determine what kind of matter fields Einstein's theory might support, for a quite general metric ansatz permitting a weak version of Birkhoff's theorem. A plethora of novel solutions is featured there, describing static hairy black holes, dressed with form fields, in the presence of suitably chosen matter actions. A particularly interesting family of solutions contained in~\cite{Bardoux:2012aw} is a flat-horizon one in which the matter content is made of multiple $k$-form fields, the number of which depends on their rank $k$ and the dimension $d$ of the transverse manifold, such that the overall stress tensor is eventually compatible with the isometries of the latter. Using these $k$-form fields (of rank not necessarily equal to $d$), the authors proceed with the construction of four- and higher-dimensional asymptotically AdS (AAdS) black holes possessing primary hair. 

Most pertinent to this work, are the four-dimensional black holes in~\cite{Bardoux:2012aw}, dubbed ``{axionic black holes}'', dressed with two constant 3-forms $H^{(i)}= p \,\dd t\wedge \dd r \wedge \dd x^i$, where $i=1,2$. By means of a Hodge duality these solutions can also be constructed using two constant 1-forms $F^{(i)}= p\, \dd x^i$. These are the field strengths of scalar fields  $\phi^{(i)}=p\, x^i$, defined up to a constant, so that any Lagrange 4-form $H^{(i)}\wedge \ast H^{(j)}\delta_{ij}$ is replaced by a sum of kinetic terms, $\dd\phi^{(i)}\wedge\ast \dd\phi^{(j)}\delta_{ij}$, which is exactly the main building block in this work. Note here, that these configurations exhibit a peculiar feature; altough the horizon is flat, the inclusion of axions leads to a lapse function with a form reminiscent of hyperbolic black holes, due to the emergence of a negative effective curvature scale, proportional to the axionic charge $p$. In the following sections, we will actually see that such a phenomenon is basically tied to the presence of standard kinetic terms with the ``correct'' sign in the action. 

Solutions with axionic charges become particularly relevant and interesting when taken as gravitational duals for certain strongly coupled quantum field theories where momentum is dissipated. In particular, in recent years, a lot of effort has been devoted to the usage of the gauge-gravity duality to understand condensed matter systems at strong coupling in which quasiparticles are absent and strong correlations are active~\cite{zaanen2015holographic,Hartnoll:2016apf,Baggioli:2019rrs}. Part of the interest has been motivated by the unusual (e.g. non-Fermi liquids like) transport properties of the so-called strange metals and the origin of superconductivity in their high-$T_c$ counterparts.
A crucial step in making this tool operative and therefore applicable to realistic condensed matter systems was to realize the fundamental importance of translational symmetry breaking and the consequent dissipation of charge carriers momentum. Abandoning the full diffeomorphism group invariance in the gravitational bulk was the key step in this direction. The explicit breaking of spatial translations of the dual field theory directly implies the presence of a finite graviton mass in the bulk~\cite{Vegh:2013sk}, and it elevates massive gravity as the universal low-energy description for condensed matter systems with finite electric conductivity~\cite{Alberte:2015isw}. Interestingly, this statement has been proven exactly, at least at the perturbative level, starting from a proper periodically modulated holographic lattice~\cite{Blake:2013owa}. 

A very convenient way to formalize theories of massive gravity is via the so-called St\"uckelberg mechanism~\cite{Hinterbichler:2011tt,Dubovsky:2004sg}. In its simplest incarnation, this formalism makes the new extra degrees of freedom explicit; they directly appear in the action in the form of a set of massless scalars whose expectation values are linearly proportional to the spacetime coordinates. In order to keep the total energy of the boundary field theory conserved, and therefore the time component of the boundary Ward identity unmodified, i.e., $\partial_a T^{at}=0$, we do not break time diffeomorphisms in the bulk. This can be ensured by considering a minimal set of scalar fields given by:
\begin{equation}
    \phi^I\,=\,p\,\delta^I_j\,x^j, \label{eq:ScalarBackgrounds}
\end{equation}
where the index $j=1,...,d$ only spans the coordinates of the transverse manifold, and $\delta^I_i$ is an isomorphism mapping to an internal space of dimension $d$ equipped with a Euclidean metric $\delta_{IJ}$. Notice that these are precisely the fields we previously discussed. We remark here that this choice is not the most general, but the only one retaining the rotational invariance in the spatial sub-manifold\footnote{See, e.g.,~\cite{Mateos:2011ix,Ge:2014aza} for examples going beyond this paradigm.}. This approach was first introduced in the applied holography community in~\cite{Andrade:2013gsa} and later generalized and understood in detail in~\cite{Baggioli:2014roa,Alberte:2015isw}. Nowadays, it is customary to call this setup ``holographic axion model''~\cite{Baggioli:2021xuv}, and to recognize it as the most versatile, simple and efficient holographic framework for breaking translational invariance. Importantly, this model is not only able to introduce momentum dissipation, but also the spontaneous (phonons dynamics)~\cite{Alberte:2017oqx} and pseudo-spontaneous~\cite{Alberte:2017cch,Ammon:2019wci} breaking mechanisms.

The big motivation to look into these models is given by the challenge to reproduce and understand the exotic transport properties of strange metals, strongly correlated materials which do not follow standard Fermi Liquids theory. Despite some initial excitement~\cite{Davison:2013txa,Blake:2014yla}, the problem still remains unsolved~\cite{Amoretti:2016cad}. 
Additionally, almost all condensed matter systems do not display relativistic invariance and most of them exhibit critical points with non-relativistic scalings (see Section 2.1.1 in~\cite{Hartnoll:2009sz}), the so-called \emph{Lifshitz scalings}:
\begin{equation}
    \Vec{x}\to\lambda\,\Vec{x}
    ,\qquad t\to\lambda^z\,t,
\end{equation}
with $z>1$~\cite{Hoyos:2010at}. Having a non-trivial Lifshitz scaling exponent seems to play an important role in the phenomenology of cuprates~\cite{PhysRevB.91.155126}. Its holographic incarnation is manifested in the so-called Lifshitz spacetime~\cite{Taylor:2015glc}, which has opened a completely new avenue for the study of strongly coupled non-relativistic field theories using the holographic duality. The whole dictionary for gravity duals with Lifshitz asymptotics has been recently formalized in~\cite{Chemissany:2014xsa} and it presents substantial differences with the standard procedure in AAdS bulk spacetimes. Given these facts, it is worth to consider more general and complicated gravitational models. In this direction, Horndeski theory~\cite{Horndeski:1974wa} is a promising framework since it constitutes a natural extension of the aforementioned holographic axion model, and it has indeed already been considered before~\cite{Baggioli:2017ojd,Figueroa:2020tya,Cisterna:2017jmv,Cisterna:2018hzf,Cisterna:2019uek,Liu:2017kml,Jiang:2017imk,Liu:2016njg,Feng:2015wvb,Feng:2015oea}.

Horndeski theory is the most general four-dimensional scalar-tensor theory with equations of motion up to second order in derivatives of the fields. Although the action contains higher-derivative self-interactions of the scalar which in general contribute higher-order terms to the field equations, it also contains gravitational \emph{counterterms}, the contribution of which results in finally having a healthy set of second-order differential equations. The shift-symmetric Horndeski action can be written as:
\begin{equation}
S=\int \sqrt{-g}\dd[4]{x}\sum_{n=2}^{5}\mathcal{L}_n,\label{eq:HornAction}
\end{equation}
in its modern generalized Galileon formulation~\cite{Deffayet:2011gz,Deffayet:2013lga,Kobayashi:2011nu}, with
\begin{align}
\mathcal{L}_2=&\,K(X)\,,\nonumber\\ \mathcal{L}_3=&-G_{3}(X)\Box\phi\,,\nonumber\\
\mathcal{L}_4=&\,G_4(X)R+G_{4X}(X)\bqty{(\Box\phi)^2-\nabla_{\mu}\nabla_{\nu}\phi\nabla^{\mu}\nabla^{\nu}\phi}\,,\nonumber\\
\mathcal{L}_5=&-\frac{G_{5X}(X)}{6}\bqty{(\Box\phi)^3-3\nabla_{\mu}\nabla_{\nu}\phi\nabla^{\mu}\nabla^{\nu}\phi\Box\phi+2\nabla^\nu\nabla_\mu\phi\nabla^\alpha\nabla_\nu\phi\nabla^\mu\nabla_\alpha\phi}\nonumber\\&+G_5(X)G_{\mu\nu}\nabla^{\nu}\nabla^{\mu}\phi\,,
\end{align}
Here, $R$ is the Ricci scalar, $G_{\mu\nu}$ is the Einstein tensor, and $G_{nX}\equiv\pdv*{G_n}{X}$ is a convenient notation. The invariance of~\eqref{eq:HornAction} under constant shifts of the scalar field, $\phi\to \phi+c$, requires that the usual dependence of the Horndeski functions $G_n$ on $\phi$ drops out; these functions only depend on the kinetic term $X\equiv -(\partial\phi)^2/2$, and thus, $\phi$ is massless. In this work, we wish to explore static planar black holes with scalar backgrounds of the form~\eqref{eq:ScalarBackgrounds} in the framework of Horndeski gravity. To do so, we have to rewrite~\eqref{eq:HornAction} so that it may support such solutions in the first place, i.e., we must promote the Horndeski scalar $\phi$ to a doublet of scalar fields $\{\phi^I\mid I=1,2\}$. Doing so, we further wish to secure the invariance of the action under both spacetime and internal transformations which means that the action itself should be a true coordinate and internal scalar with no free indices. Internal transformations here will actually be appropriate scalar field reparametrizations, as we will later see in detail. Clearly, this can only happen if we discard $\mathcal{L}_3$ and $\mathcal{L}_5$, or equivalently, in terms of the original Horndeski action~\eqref{eq:HornAction}, if we only keep the sector invariant under reflection transformations $\phi\to -\phi$. The explicit form of the truncated action for the doublet will be given in the next section, and we will refer to it as \emph{axionic Horndeski theory}. 

Note here that one could judiciously question the necessity of forming internal scalars in order to obtain bulk solutions with scalar backgrounds as in~\eqref{eq:ScalarBackgrounds}. Indeed, we might as well introduce two scalar fields $\phi^I$ by just adding two copies of~\eqref{eq:HornAction}, one for each scalar\footnote{It is not clear how this strategy could be reconciled with the standard massive gravity language in terms of St\"uckelberg fields~\cite{Dubovsky:2004sg} and with the common EFT formalism~\cite{Nicolis:2015sra}.}. However, it is interesting, that even in this case, in order for~\eqref{eq:HornAction} to support solutions with scalar backgrounds linear in the coordinates of the transverse manifold, one needs to perform a significant truncation, namely, $\mathcal{L}_3$ and $\mathcal{L}_5$ must again be discarded, but for a different reason now. The reason has to do with the equations of motion for the scalar fields, which are just on-shell closure statements for the two Noether currents associated with the aforementioned shift symmetry. By direct inspection, one can verify that for a profile~\eqref{eq:ScalarBackgrounds}, the equation of motion for the $I$th scalar field cannot be satisfied essentially due to the presence of a radially dependent radial component of the $I$th current. It can be further shown that this radial component is solely a variational artefact of having $\mathcal{L}_3$ and $\mathcal{L}_5$ in the action; once we discard those, the $I$th current does only possess a radially dependent component in the $I$th direction of the transverse manifold, and therefore, the corresponding scalar equation of motion is identically satisfied. Nevertheless, we won't be dealing with such a prescription, and we just mention the above for completeness.  

Finally, the main goal of this work is to establish a systematic approach to hairy planar black holes in axionic Horndeski theory. In simpler words, if such exact solutions exist, we wish to be able to derive them with minimal effort by just specifying the Horndeski functions, and executing some straightforward process steps. In this sense, we want to construct a step sequence which results in solutions once executed, a \emph{solution-generating algorithm} if you wish. Note that although we only display certain examples, this algorithm can in general produce many more exact solutions with the aforementioned characteristics, and in that fashion, the principal result of this manuscript is a rather general one. Based on the logic behind this prescription, and armed only with a minimal set of initial assumptions, we will also be able to pull off a complete derivation of the DC transport matrix for relativistic field theory (holographic) duals, without the need to fix the $G_n$'s a priori. Although this part will not be fully developed in terms of an exhaustive analysis, this is nevertheless a potentially useful result, as universal linear-response features might possibly be extracted from it.

\subsubsection*{Plan of this work.}
The manuscript is organized as follows: in section \ref{sec:Theory} we present our general setup and its main features; in section \ref{sec:Algo} we present the novel algorithm which can be used to generate planar black hole solutions with axionic hair, this being the main result of our work; in section \ref{sec:Examples} we provide two examples of how this algorithm works;
in section \ref{sec5} we initiate the study of the relativistic field theory duals of these solutions by deriving their DC transport properties; finally, in section \ref{sec6} we conclude and discuss possible future directions.

\section{The general setup: theory and field equations}\label{sec:Theory}
 
 In this work, we deform the Einstein-Maxwell (EM) action by adding to it a Horndeski action with reflection and shift symmetry. Instead of a single scalar field, we introduce $d$-many axionic scalar fields, or simply axions, in general $d+2$ spacetime dimensions, packed in a $d$-tuple $\phi=(\phi^1,...,\phi^d)$ which transforms as a vector under the action of an internal group $O(d)$. This ``flavor'' vector can be used to construct a scalar quantity $\Bar{X}$:
\begin{equation}
    \Bar{X}=\mathrm{Tr}\,\mathcal{I}=-\frac{1}{2}g^{\mu\nu}\partial_\mu\phi^I\partial_\nu \phi^J\delta_{IJ},\label{eq:KineticTrace}
\end{equation}
where $\mathcal{I}^{IJ}\equiv -\frac{1}{2}g^{\mu\nu}\partial_\mu\phi^I\partial_\nu \phi^J$, capital Latin indices $I,J,...$ are internal-space indices ranging from 1 to $d$, and the trace is taken with respect to the reference metric $\delta_{IJ}=(1,1,...,1)_{IJ}$, which is the canonical static metric in $d$-dimensional Euclidean internal space.\footnote{Einstein's summation convention is adopted also for internal-space indices.} This is basically the main building block of our theory, as all Horndeski functions will be functions of~\eqref{eq:KineticTrace}. Notice that for $d>1$, this is not the most exhaustive choice; already for $d=2$, there is another algebraically independent contraction:
\begin{equation}
    \det \mathcal{I}=\frac{1}{2}\pqty{\Tr \mathcal{I}\Tr \mathcal{I}-\Tr \mathcal{I}^2}.
\end{equation}
Indeed, for a $d$-dimensional transverse manifold, $\mathcal{I}$ is a $d\times d$ matrix, and $\Tr\mathcal{I}^n$ with $n\leq d$ are independent scalar objects which can, and in general have to, appear in the action via appropriate combinations.\footnote{See for example~\cite{Nicolis:2013lma,Alberte:2015isw,Alberte:2018doe}.} Nevertheless, in this work we do not consider other combinations.

More concretely, we choose to work with the following action principle 
\begin{equation}
    S\equiv S_{EM}+S_{AH}\label{eq:TotalAction}\,,
\end{equation}
where 
\begin{eqnarray}
S_{EM}&=&\int \sqrt{-g}\dd[d+2]{x}\pqty{R-2\Lambda -\frac{1}{4}F_{\mu\nu}F^{\mu\nu}},\label{eq:EMAction}\\
S_{AH}&=&\int \sqrt{-g}\dd[d+2]{x}\bqty{K(\Bar{X})+G_4(\Bar{X})R+G_{4\Bar{X}}(\Bar{X})\delta^{\rho\sigma}_{\mu\nu}\phi^{I\mu}_\rho \phi^{J\nu}_\sigma \delta_{IJ}}.\label{eq:HorndeskiAction}
\end{eqnarray}
Here, we have set the coupling in front of the Einstein-Hilbert piece in~\eqref{eq:EMAction} to unity which amounts to setting $M_{Pl}^d=16\pi$, together with natural units. Moreover, $\Lambda$ denotes the bulk cosmological constant, and the tensor $F_{\mu\nu}=\nabla_\mu A_\nu - \nabla_\nu A_\mu$ stands for the field strength of a bulk $U(1)$ gauge field $A_\mu$. We also use the convenient notation:
\begin{equation}
    \phi^{I\mu\nu...}{}_{\rho\sigma}\equiv \nabla^\mu\nabla^\nu...\nabla_\rho\nabla_\sigma...\phi^I,
\end{equation}
where the internal index should always precede the coordinate ones. Finally, the last piece in~\eqref{eq:HorndeskiAction} can be cast into the more familiar form:
\begin{equation}
    \delta^{\rho\sigma}_{\mu\nu}\phi^{I\mu}_\rho \phi^{J\nu}_\sigma \delta_{IJ}=\Box\phi^I\Box\phi_I-\phi^I_{\mu\nu}\phi_{I\mu\nu},
\end{equation}
and one may then immediately recognize the quartic sector of~\eqref{eq:HornAction}.

Let us first discuss the symmetries of~\eqref{eq:TotalAction}. Apart from being a coordinate scalar, the action is also an internal-space scalar. Therefore, it is invariant under general coordinate transformations, but also under scalar-field reparametrizations which are isometries of the Euclidean plane. The latter are transformations
\begin{equation}
    \phi^I\to \psi^I=\Lambda^I{}_J\phi^J+c^J,
\end{equation}
where $\Lambda$ is a static matrix in $O(d)$ and $c$ is a constant shift vector. As mentioned in the introduction, the action~\eqref{eq:HorndeskiAction} is the only sector of the $(d+2)$-dimensional version of~\eqref{eq:HornAction}, which, after promoting the Galileon to a $d$-tuple, enjoys these symmetries. Indeed, in the other sectors, $\phi$ appears an odd number of times which implies that these coordinate scalars would transform as $O(d)$ vectors once we replace the scalar field by the flavor vector. For example, the higher-derivative self-interaction terms in $\mathcal{L}_5$ would in general have three free internal indices, but only two of them could be contracted with the reference metric, thus leaving us with one free internal index. Additionally, note here that once a background configuration is chosen for the scalar fields, the full symmetry will be broken down to a diagonal subgroup of internal and spacetime symmetries (see~\cite{Alberte:2015isw} for more details).

To extract the bulk equations of motion, we vary~\eqref{eq:TotalAction} with respect to the metric, the gauge field and the axions $\phi^I$. Variation with respect to the axions, using
\begin{equation}
    \var \Bar{X}=-\phi_I^\mu \var \phi^I_\mu,
\end{equation}
yields $\partial_\mu(\sqrt{-g}J^\mu_I)=0$, where 
\begin{eqnarray}
J^\mu_I&\equiv& -\pqty{K_{\Bar{X}}+G_{4\Bar{X}}R+G_{4\Bar{X}\Bar{X}}\delta^{\rho\sigma}_{\lambda\nu}\phi^{J\lambda}_\rho \phi^\nu_{J\sigma}}\phi_I^\mu-2\nabla^\rho\pqty{G_{4\Bar{X}}\phi^\sigma_{I\nu}\delta^{\mu\nu}_{\rho\sigma}}\nonumber\\
&=&2G_{4\Bar{X}\Bar{X}}\phi_{J\nu}\pqty{\phi^{J\mu\nu}\Box \phi_I - \phi^{J\nu\rho}\phi^\mu_{I\rho}}+2G_{4\Bar{X}}G^{\mu\nu}\phi_{I\nu}\nonumber\\
&&-\pqty{K_{\Bar{X}}+G_{4\Bar{X}\Bar{X}}\delta^{\rho\sigma}_{\lambda\nu}\phi^{J\lambda}_\rho \phi^\nu_{J\sigma}}\phi_I^\mu,\label{eq:NoetherCurrentI}
\end{eqnarray}
represents the $I$th Noether current associated with the constant shift $\phi^I\to \phi^I + c^I$. Hence, the equations of motion for the $d$-many axions are just (on-shell) closure statements for the $d$-many such currents. Moving on, the gauge-field equations of motion are the standard ones: 
\begin{equation}
    \partial_\nu\pqty{\sqrt{-g}F^{\nu\mu}}=0\,.
\end{equation}
Finally, the modified Einstein equations, obtained by varying~\eqref{eq:TotalAction} with respect to the metric, are 
\begin{equation}
    \mathcal{E}_{\mu\nu}\equiv G_{\mu\nu}+g_{\mu\nu}\Lambda -\frac{1}{2}\pqty{F_{\mu\rho}F_\nu{}^\rho-\frac{1}{4}g_{\mu\nu}F^{\rho\lambda}F_{\rho\lambda}}-T_{\mu\nu}^{(\phi)}=0,\label{eq:EinsteinEquations}
\end{equation}
where
\begin{eqnarray}
    T^{(\phi)}_{\mu\nu}&=&\frac{1}{2}\pqty{K_{\Bar{X}}\phi^I_\mu \phi_{I\nu}+Kg_{\mu\nu}}-G_{4}G_{\mu\nu}+G_{4\Bar{X}}\pqty{\phi^I_{\mu\nu}\Box\phi_I-\phi^I_{\rho\mu}\phi^{\rho}_{I\nu}}\nonumber\\
    &+&G_{4\Bar{X}}\pqty{\frac{1}{2}R\phi^I_\mu  \phi_{I\nu}+g_{\mu\nu} R^{\rho\lambda}\phi^I_\rho  \phi_{I\lambda}-2 R_{\lambda(\mu}\phi^I_{\nu)} \phi^\lambda_I-R_{\mu\rho\nu\lambda}\phi^{I\rho} \phi_I^\lambda}\nonumber\\
    &+&\frac{1}{2}\pqty{\Box \phi^I \Box \phi_I -\phi^I_{\rho\sigma} \phi_I^{\rho\sigma}}\pqty{G_{4\Bar{X}\Bar{X}}\phi^J_\mu  \phi_{J\nu} - G_{4\Bar{X}}g_{\mu\nu}}\nonumber\\
    &+&G_{4\Bar{X}\Bar{X}}\bqty{(\phi^{I\rho} \phi_{I\rho\mu})(\phi^{J\sigma} \phi_{J\sigma\nu})-g_{\mu\nu}(\phi^I_{\rho\lambda} \phi_I^\rho)(\phi_\sigma^{J\lambda}  \phi^\sigma_J)}\nonumber\\
    &+& G_{4\Bar{X}\Bar{X}}\bqty{\phi^I_{\lambda\rho} \phi^\rho_I\pqty{2\phi^{J\lambda}_{(\mu}  \phi_{J\nu)}-\phi^{J\lambda} \phi_{J\mu\nu}}+(\phi^I_{\rho\lambda} \phi_I^\rho)\Box \phi^J \pqty{g_{\mu\nu}\phi_J^\lambda-2\delta^\lambda_{(\mu}\phi_{|J|\nu)}}}.\nonumber\\
\end{eqnarray}
To get the above, we also used the fact that for any scalar function $G(\Bar{X})$, it holds that 
\begin{equation}
    \nabla_\mu G=G_{\Bar{X}}\nabla_\mu \Bar{X}=-G_{\Bar{X}}\phi^I_{\mu\nu} \phi^\nu_I\,.
\end{equation}

\section{A solution-generating algorithm}\label{sec:Algo}

The main task of this work is to construct a systematic way of obtaining all solutions following from~\eqref{eq:TotalAction}, which further satisfy the following requirements:
\begin{enumerate}[(i)]
    \item The bulk geometry is four-dimensional, represented by the static ansatz:
    \begin{equation}
        \dd s^2 =g_{\mu\nu}\dd x^\mu \dd x^\nu= -h(r) \dd t^2 +\frac{\dd r^2}{f(r)} + r^2 \delta_{ij}\dd x^i \dd x^j,\qquad i,j=1,2,\label{eq:MetricAnsatz}
    \end{equation}
    where $h,f$ are to be found by solving the bulk equations of motion. The coordinates $x^i$ of the transverse manifold are considered noncompact, and in this regard, the spacetime is truly plane-symmetric. The boundary is located at radial infinity, $r=\infty$.
    \item The gauge potential is a function of only the radial coordinate $r$ with the only non-vanishing component being the temporal one, $A_t(r) \neq 0$. In particular, the gauge field reads $A_\mu=(a(r),0,0,0)$.
    \item As already anticipated in \eqref{eq:ScalarBackgrounds}, the bulk solution for the axions is $\phi^I=p\, \delta^I_i x^i$.
\end{enumerate}
With these requirements, one can also straightforwardly verify that $\mathcal{E}_{\mu\nu}$ is compatible with the symmetries of~\eqref{eq:MetricAnsatz}, i.e., $\mathsterling_\xi\mathcal{E}_{\mu\nu}=0$ where the Lie derivative is taken with respect to all the corresponding Killing vectors. Note, however, that because of~\eqref{eq:ScalarBackgrounds}, $\mathsterling_\xi \phi^I\neq 0$, and in this sense, the \emph{full} solution (that is, including the gauge and scalar backgrounds) is only static. Using the algorithm, we will be able to generate plane-symmetric solutions to the bulk equations of motion, but we won't be able to tell if these describe planar black holes. This is an issue that has to be treated separately for each case. We will put the algorithm to use, providing novel closed-form hairy black holes which we classify according to the form of $G_4$. We will refer to these  black hole solutions as \emph{axionic black branes} \cite{Caldarelli:2016nni}. Moreover, in this work we will not make statements about the physical permissibility of these configurations in terms of instabilities, energy conditions etc. In this vast landscape of solutions, it is reasonable to expect the existence of pathological setups which should be avoided.

Given these premises, we have
\begin{equation}
    \Bar{X}=-\frac{p^2}{r^2},\qquad F_{\mu\nu}=-a'(r)\delta^{tr}_{\mu\nu},\label{eq:KineticTermAnsatz}
\end{equation}
where $'\equiv \pdv*{}{r}$. Here, we can make a first observation which will prove to be crucial for the success of the algorithm. The first equation in~\eqref{eq:KineticTermAnsatz} can be inverted in order to find
\begin{equation}
    r(\Bar{X})=\frac{\abs{p}}{\sqrt{-\Bar{X}}},\label{eq:rofX}
\end{equation}
with $\Bar{X}<0$ by definition. As a consequence of the reflection symmetry, $p$ will always appear inside a modulus, and henceforth, we take $p>0$ without loss of generality. It turns out that if we treat $\Bar{X}$ as a variable, instead of the radial coordinate $r$, the equations of motion become much more tractable, and an algorithmic approach can be established. Radial derivatives of any function $G(r)$ can be written as $\Bar{X}$ derivatives, e.g., 
\begin{eqnarray}
    G'(r)&=&\frac{2(-\Bar{X})^{3/2}}{p}G_{\Bar{X}},\nonumber\\
    G''(r)&=&-\frac{2\Bar{X}^2}{p^2}\pqty{3G_{\Bar{X}}+2\Bar{X}G_{\Bar{X}\Bar{X}}},
\end{eqnarray}
and we will use this fact, together with~\eqref{eq:rofX}, to transform ordinary differential equations (ODEs) for $r$ to ODEs for $\Bar{X}$.

Let us first start with the equations of motion for the axions. In the previous section, we saw that the $I$th axion must satisfy $\partial_\mu(\sqrt{-g}J_I^\mu)=0$. To verify that $\phi^I=p\,\delta^I_i x^i$ is a solution, we repeat here the argument made in the introduction of this work. Evaluating~\eqref{eq:NoetherCurrentI}, one finds that 
\begin{equation}
    J^\mu_I=j(r)\delta^\mu_i\delta^i_I,
\end{equation}
where $j(r)$ is some complicated function of the radial coordinate. Therefore, it immediately follows that
\begin{equation}
    \partial_\mu(\sqrt{-g}J^{\mu}_I)=\partial_{x^i}\pqty{\sqrt{-g(r)}j(r)\delta^i_I}=0,
\end{equation}
identically. Next, the equations of motion for the bulk gauge field can be solved for a closed-form expression of the electric field:
\begin{equation}
    a'(r)=\sqrt{\frac{h}{f}}\frac{q_e}{r^2},\label{eq:ElectricField}
\end{equation}
where the integration constant $q_e$ is related to the charge density in the dual field theory. We note here that one may also include a magnetic charge by considering a dyonic profile for $A_\mu$. Since the electromagnetic sector of the action consists of just the standard Maxwell term, at least at the level of the background solutions, the presence of magnetic components results only in an additive contribution; that is to say, solutions with a dyon can be obtained from the electrically charged solutions we present, by sending $q_e^2 \rightarrow q_e^2+q_m^2$. As expected, dyonic solutions will be invariant under the duality exchange $q_e\leftrightarrow q_m$. Although this may be not that interesting from a gravity point of view, considering dyons opens several new possibilities for the transport properties of the dual field theory. We leave the study of the dyonic solutions for the near future.

Proceeding with the modified Einstein equations~\eqref{eq:EinsteinEquations}, it is possible to show that, given the solutions $\phi^I=p\,\delta^I_i x^i$ and~\eqref{eq:ElectricField}, the $ii$ components are not independent, and therefore we only need to satisfy $\mathcal{E}_{tt}=0=\mathcal{E}_{rr}$ in order to solve the full system. 
This has been proven in full generality in \cite{Hervik:2019gly,Hervik:2020nxs,Hervik:2020zvn}, where the authors show how to generalize a Schwarzschild-like ansatz in order to construct static black hole solutions for any Lagrangian density (describing a metric theory of gravity) being a functional of the Riemann tensor and its derivatives of any order, arbitrary $k$-form fields and scalar fields. The metric ansatz is generalized by considering an isotropy-irreducible homogeneous base space which allows, before specifying the particular theory under scrutiny, to ensure that solutions of the system can be found by solving two ODEs, one for $g_{tt}$, and the other for $g_{rr}$ (please refer to~\cite{Hervik:2020zvn} for further details).

The $tt$ component yields a first-order inhomogeneous ODE for $f$ with function coefficients $\alpha_n$. In the $\Bar{X}$ formalism, it reads
\begin{equation}
    \alpha_0+\alpha_1 f+ \alpha_2 f_{\Bar{X}}=0,\label{eq:fODE}
\end{equation}
with 
\begin{eqnarray}
    \alpha_0&\equiv&-\Lambda + \frac{K}{2}-\frac{q_e^2 \Bar{X}^2}{4 p^4},\\
    \alpha_1&\equiv&\frac{\Bar{X}}{p^2}\pqty{1+G_4+4 \Bar{X}^2 G_{4\Bar{X}\Bar{X}}},\\
    \alpha_2&\equiv&-\frac{2\Bar{X}^2}{p^2}\pqty{1+G_4-\Bar{X}G_{4\Bar{X}}}.
\end{eqnarray}
Then, $\mathcal{E}_{rr}$ yields another ODE, for $h$ this time, which acquires the particularly simple form
\begin{eqnarray}
\frac{h_{\Bar{X}}}{h}=-\frac{\alpha_3}{\alpha_4},\label{eq:hODE}
\end{eqnarray}
with
\begin{eqnarray}
    \alpha_3&\equiv&-\frac{4\alpha_0 p^2}{f}- 4\Bar{X}(1+G_4-2\Bar{X}G_{4X}),\qquad \alpha_4\equiv-4 p^2 a_2.
\end{eqnarray}
The general solution to~\eqref{eq:fODE} is obtained via the integrating-factor method:
\begin{equation}
    f(\Bar{X})=\exp({-\int^{\Bar{X}}\dd u\frac{\alpha_{1}(u)}{\alpha_{2}(u)}})\bqty{f_0-\int^{\Bar{X}}\dd u \exp({\int^u\dd v\frac{\alpha_{1}(v)}{\alpha_{2}(v)}})\frac{\alpha_{0}(u)}{\alpha_{2}(u)}},\label{eq:GeneralSolutionf}
\end{equation}
provided that $\alpha_2\neq 0$, whereas~\eqref{eq:hODE} can be solved for 
\begin{equation}
    h(\Bar{X})=h_0\exp({-\int^{\Bar{X}}\dd u\frac{\alpha_{3}(u)}{\alpha_{4}(u)}})\label{eq:GeneralSolutionh}.
\end{equation}
Here, both $f_0$ and $h_0$ are integration constants.

To simplify things, we now set $h(r)=C(r)f(r)$, and observe that a master equation for $C$ can be found. It is given by $\mathcal{E}_{tt}h^{-1}+\mathcal{E}_{rr}f=0$, reading 
\begin{equation}
    \frac{C_{\Bar{X}}}{C}=-\frac{G_{4\Bar{X}}+2\Bar{X}G_{4\Bar{X}\Bar{X}}}{1+G_4-\Bar{X}G_{4\Bar{X}}}.\label{eq:MasterEquationC}
\end{equation}
Remarkably, $C$ knows nothing of k-essence, and it can be directly integrated from the above equation once $G_4$ is chosen. Of course, a closed form result depends on whether the $\Bar{X}$ integral of the right-hand side (rhs) of~\eqref{eq:MasterEquationC} has a closed form expression, or not. Note here that the denominator of~\eqref{eq:MasterEquationC} is $\propto \alpha_2$, and thus, nonzero. Another interesting remark can be made. To obtain solutions with $g^{rr}=-g_{tt}$, the rhs of the above equation must vanish, and this can only happen if $G_4$ is constant, or for $G_4 \propto \sqrt{-\Bar{X}}$. The general black brane solution for this particular model has been already found in previous work~\cite{Figueroa:2020tya}, but with a slightly different action and building blocks.\footnote{Instead of a flavor vector and a trace $\Bar{X}$, $d$-many axions were introduced, each via a copy of the Horndeski action, with the building blocks being internal-space vectors $X^I$, namely, the diagonal elements of what we here defined as $\mathcal{I}$.} Now that we know $C$, we can find a better form for $f$ by massaging~\eqref{eq:GeneralSolutionf} in order to arrive at
\begin{equation}
    f=\frac{2(-\Bar{X})^{5/2}}{\alpha_2\sqrt{C}p^2}\pqty{f_0-k+\int^{\Bar{X}}\dd u\frac{\sqrt{C(u)}\pqty{q_e^2u^2+4\Lambda p^4}}{8p^2(-u)^{5/2}}},\label{eq:BetterFormf}
\end{equation}
with
\begin{equation}
    k\equiv \int^{\Bar{X}}\dd u\frac{\sqrt{C(u)}K(u)p^2}{4(-u)^{5/2}}.\label{eq:kEquation}
\end{equation}
The above form is much more tractable now, and we will see that with~\eqref{eq:BetterFormf} and~\eqref{eq:MasterEquationC} at hand, we can rapidly obtain solutions which would otherwise be hard to directly integrate from the field equations.

Let us now assume that the solution describes a black brane with its event horizon located at $r=r_0$. Via the standard Wick-rotation method and the avoidance of the conical singularity in the Euclidean section, we can find the Hawking temperature which in the $\Bar{X}$ formalism assumes the form:
\begin{equation}
    T_H=\frac{(-\Bar{X})^{3/2}\sqrt{C}\abs{f_{\Bar{X}}}}{2 \pi p}\mid_{\Bar{X}=\Bar{X}_0}=\frac{(-\Bar{X})^{3/2}\sqrt{C}}{2 \pi p}\abs{\frac{\alpha_0}{\alpha_2}}\mid_{\Bar{X}=\Bar{X}_0},
\end{equation}
with $\Bar{X}_0\equiv \Bar{X}(r_0)$. In a less obscure format, in terms of $r_0$,
\begin{equation}
    T_H=\frac{\sqrt{C}}{16\pi r_0}\abs{\frac{q_e^2+2r_0^4(2\Lambda-K)}{r_0^2(1+G_4)+G_{4\Bar{X}}p^2}},\label{eq:HawkingTemperature}
\end{equation}
where all functions are evaluated at $r_0$. Notice that $C(r_0)$ must be positive for $T_H$ to be real; needless to say, we must choose $C(r)>0$ for all $r\geq r_0$ in order to have positive $-g_{tt},g^{rr}$ in the exterior. 

At this early stage, we can also extract an expression for the Wald entropy \cite{Wald:1993nt}:
\begin{equation}
    s_W=-2\pi \int_\Sigma \dd[2]x r_0^2 \pdv{\mathcal{L}}{R_{\mu\nu\rho\sigma}}\hat{\epsilon}_{\mu\nu}\hat{\epsilon}_{\rho\sigma},\label{eq:WaldEntropyGeneral}
\end{equation}
where $\hat{\epsilon}_{\mu\nu}$ is the unit binormal to the bifurcation surface $\Sigma$ with normalization $\hat{\epsilon}_{\mu\nu}\hat{\epsilon}^{\mu\nu}=-2$, and 
\begin{equation}
    \pdv{\mathcal{L}}{R_{\mu\nu\rho\sigma}}=(1+G_4)g^{\rho[\mu} g^{\nu]\sigma}+G_{4\bar{X}}\delta^{[\mu}_\alpha\delta^{\nu]}_\beta \delta^{[\rho}_\gamma \delta^{\sigma]}_\delta g^{\alpha\gamma}\phi^\beta \phi^\delta,
\end{equation}
after integrating~\eqref{eq:HorndeskiAction} by parts. Notice that due to the particular axion backgrounds, the entropy obtained by computing~\eqref{eq:WaldEntropyGeneral} after integrating~\eqref{eq:TotalAction} by parts does exactly agree with the result obtained by directly calculating~\eqref{eq:WaldEntropyGeneral} for~\eqref{eq:TotalAction}. The explicit expression reads:
\begin{equation}
    s_W=4\pi\omega r_0^2\pqty{1+G_4(\Bar{X}_0)},\qquad \omega\equiv \int_\Sigma \dd[2]x,\label{eq:WaldEntropy}
\end{equation}
and we see that the result is completely ignorant of k-essence since the latter couples minimally to gravity. Note that one can define an effective running gravitational constant, in particular,
\begin{equation}
    G^{-1}_{eff}(r)\equiv M_{Pl}^2+16\pi G_4,
\end{equation}
thereby re-expressing~\eqref{eq:WaldEntropy} as
\begin{equation}
    s_W=\frac{\mathcal{A}_0}{4 G_{eff}(r_0)}, \qquad \mathcal{A}_0=\omega r_0^2.
\end{equation}
In units $G_{eff}(r_0)=1$ the renown ``quarter law'' is then manifest \cite{Brustein:2007jj}. However, this definition must come with $G_4> 0$, otherwise $G_{eff}$ will become negative at some $r$. Although we will not define such an effective object, we will take $G_4$ to be positive in order to avoid imposing more delicate conditions on $r_0$. Finally, bear in mind that the plane extends to infinity, and thus, for $-\ell<x^i<\ell$ with $\ell \to \infty$,
\begin{equation}
    \omega=\lim_{\ell \to \infty} \int_{-\ell}^\ell \dd x^1\int_{-\ell}^\ell \dd x^2=\lim_{\ell\to \infty} 4\ell^2.
\end{equation}
The quantity of interest will therefore be the entropy density:
\begin{equation}
    \tilde{s}=4\pi r_0^2(1+G_4(\bar{X}_0)),
\end{equation}
and by ``finite entropy'' in the following text we will mean a finite $\tilde{s}$.

Since we know the Hawking temperature, we can take things one step further, and assume the existence of extremal black holes, zero-temperature black hole solutions to the bulk equations of motion. Given a bulk solution at Hawking temperature $T_H$, its extremal version, if it exists, can be found by solving $T_H=0$ for $r_e$, the largest positive real root of $T_H$ which also is a double zero of $h$ and $f$. In equivalent words, we need to find the largest solution to 
\begin{equation}
    -q_e^2-4\Lambda r^4 +2K r^4=0.\label{eq:ExtremalPoly}
\end{equation}
It is interesting that $G_4$ does not enter in the above, and we can conclude that extremal branes come with sizes independent of the quartic-sector contributions, although the latter are expected to affect the mass of the extremal configurations. Before closing this section, we would like to see what information we can get from the zero-temperature Einstein equations without knowing the explicit forms of $h,f$ that solve the field equations in the whole bulk. To that end, let us work with the near-horizon approximation $f(r)=f''(r_e)(\epsilon z)^2/2 +\mathcal{O}(\epsilon^3)$ and similarly for $h$, where we defined $\epsilon z\equiv r-r_e$. Given an electric field~\eqref{eq:ElectricField} and the axion backgrounds~\eqref{eq:ScalarBackgrounds}, we do a series expansion of the metric-field equations about $\epsilon=0$ with $z$ held fixed. To leading order we find that the $tt$ and $rr$ components are simply the statement $T_H(r_e)=0$. However, nontrivial constraints are coming from the $ii$ components, in particular, 
\begin{equation}
    f''(r_e)=\frac{q_e^2+K_{\Bar{X}}p^2r_e^2}{r_e^4(1+G_4)+G_{4\Bar{X}}p^2 r_e^2},\label{eq:zeroTempiiConstraint}
\end{equation}
all functions evaluated at $r_e$.

Additionally,~\eqref{eq:MetricAnsatz} becomes 
\begin{equation}
    \dd s^2\sim -\frac{h''(r_e)(\epsilon z)^2\dd t^2}{2}+\frac{2\dd z^2}{f''(r_e)z^2}+(r_e+\epsilon z)^2\delta_{ij}\dd x^i\dd x^j,
\end{equation}
where $h''(r_e)=C(r_e)f''(r_e)$. We now need to define a new time coordinate $\tau$ in order to absorb $\epsilon$ so that we can then send $\epsilon\to 0$ with $\tau,z$ held fixed obtaining the extremal IR metric. Since $C(r_e)$ is ultimately a number, it can also be absorbed into $\tau$ by $t\to \tau/(\sqrt{C(r_e)}\epsilon)$. Combining this with the optional reparametrization $u(z)\equiv2/(f''(r_e)z)$, leads us to the near-horizon $AdS_2\times\mathbb{R}^2$ form:
\begin{equation}
    \dd s_{IR}^2\underset{\epsilon\to 0}{=}\frac{L_{IR}^2}{u^2}\pqty{-\dd \tau^2 +\dd u^2}+r_e^2\delta_{ij}\dd x^i\dd x^j,\label{eq:IRmetric}
\end{equation}
where we have identified 
\begin{equation}
    L_{IR}^2=\frac{2}{f''(r_e)}.\label{eq:AdS2size}
\end{equation}
The above makes sense for $f''(r_e)>0$. Note that at this stage, this is only a solution in the deep infra-red. However, once we find a bulk solution with a zero-temperature configuration, the extremal setup will automatically describe a geometry interpolating between~\eqref{eq:IRmetric} in the far IR and some solution-dependent asymptotic geometry in the UV.\footnote{UV and IR do of course make sense only in the language of the dual field theory. That being the case, the domain wall picture is, roughly speaking, a ``geometrization'' of the RG flow in the dual system.} Finally notice from~\eqref{eq:zeroTempiiConstraint} that the extremal geometry exists also in the absence of an electric charge, purely supported by the axionic parameter $p$, provided that $K_{\bar{X}}\neq 0$.

\subsubsection*{The ``user-friendly'' algorithm }

\begin{figure}
    \centering
    \includegraphics{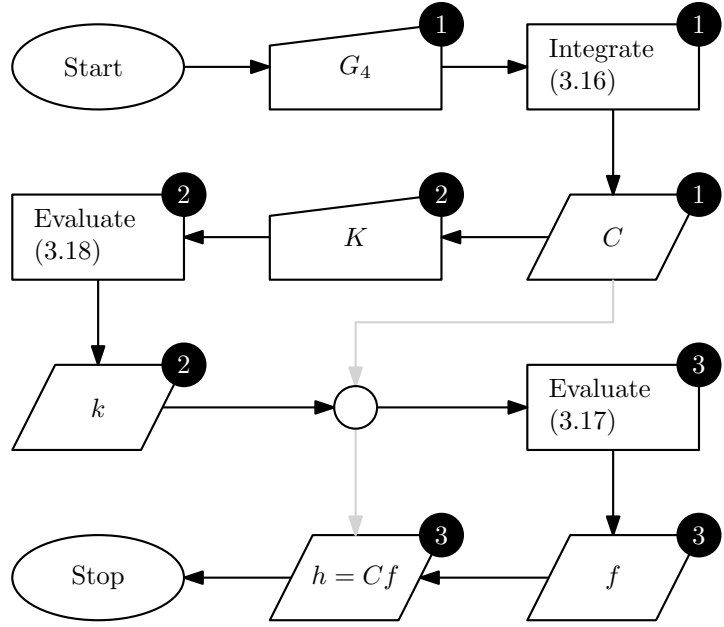}
    \caption{Flowchart description of the algorithm. We can group the steps into three phases which are indicated by numbers enclosed in disks. Ellipses indicate start/stop, trapeziums are for manual input, rectangles denote process steps, parallelograms are for output and the circle is a combination of outputs from phases 1 and 2.}
    \label{fig:FlowchartFig}
\end{figure}

To conclude, let us give a ``user-friendly'' version of what we just presented with the visual aid of the flowchart in figure~\ref{fig:FlowchartFig}. Given the solutions~\eqref{eq:ElectricField} and $\phi^I=p\,\delta^I_ix^i$, we can group the total process of solving the modified Einstein equations into three phases:
\begin{itemize}
    \item{\textbf{Phase 1.}} 
    \begin{enumerate}[(i)]
        \item Choose $G_4$ in \eqref{eq:HorndeskiAction}.
        \item Integrate the rhs of~\eqref{eq:MasterEquationC} with respect to $\Bar{X}$, then exponentiate to obtain $C$.
    \end{enumerate}
    This results in an expression for $C$.
    \item \textbf{Phase 2.} 
    \begin{enumerate}[(i)]
        \item Choose $K$ in \eqref{eq:HorndeskiAction}.
        \item Evaluate the integral in the rhs of~\eqref{eq:kEquation} to obtain $k$.
    \end{enumerate}
    This results in an expression for $k$.
    \item \textbf{Phase 3.} 
    \begin{enumerate}[(i)]
        \item Collect outputs from previous phases.
        \item Evaluate the expression in the rhs of~\eqref{eq:BetterFormf} to obtain $f$.
        \item Do $Cf$ to obtain $h$.
    \end{enumerate}
    This results in expressions for $f$ and $h$.
\end{itemize}
Note here that phase 2 can be skipped, and the solution can be given in terms of $k$. In what follows, we will initially skip phase 2 in order to do a preliminary analysis of the solution, but we will always complete this phase afterwards. In this sense, solutions should be first classified with respect to $G_4$.

Further data can be also collected by direct evaluation:
\begin{itemize}
    \item Choose $G_4$ and $K$, complete phase 1, and evaluate the rhs of~\eqref{eq:HawkingTemperature} to obtain $T_H$. 
    \item Choose $G_4$ and evaluate the rhs of~\eqref{eq:WaldEntropy} to obtain $s_W$.
    \item Choose $K$ and solve~\eqref{eq:ExtremalPoly} for $r_e$ (if applicable).
    \item Choose $K$ and $G_4$, determine $r_e$ from previous process, evaluate the rhs of~\eqref{eq:zeroTempiiConstraint}, and substitute it into~\eqref{eq:AdS2size} to find the size of the fixed-point $AdS_2$ (if applicable).
\end{itemize}
In the next section, we initiate a first exploration of the vast landscape of models captured by~\eqref{eq:TotalAction}, with the assistance of the algorithm. 

\section{Novel black branes in Hordenski theory}\label{sec:Examples}

\subsection{\boldmath The $G_4\propto \bar{X}$ class}

As our first example, we set $G_4=\gamma\Bar{X}$. Completing phases 1 and 3, we quickly arrive at the full solution:
\begin{eqnarray}
    f&=&-\frac{\Lambda r^2}{3}-\exp(-\frac{\hat{\gamma}}{2 r^2})\frac{M-pk}{r}-\frac{\hat{\gamma} \Lambda}{3} +\frac{3q_e^2+4\Lambda\Hat{\gamma}^2}{6r\sqrt{2\hat\gamma}}{D}\pqty{\frac{\sqrt{\Hat{\gamma}}}{r\sqrt{2}}},\\
    C&=&\exp(\frac{\hat{\gamma}}{r^2}),\qquad a=\mu -\sqrt{\frac{\pi}{2\Hat{\gamma}}}q_e\mathrm{erfi}\pqty{\sqrt{\frac{\hat{\gamma}}{2}}\frac{1}{r}}, \qquad \phi^I=p\delta^I_ix^i.\label{eq:FullsolutionG4propX}
\end{eqnarray}
Here, $\mu$ is the chemical potential in the dual field theory, given in terms of horizon data so that $a(r_0)=0$, i.e., 
\begin{equation}
    \mu=\sqrt{\frac{\pi}{2\Hat{\gamma}}}q_e\mathrm{erfi}\pqty{\sqrt{\frac{\Hat{\gamma}}{2}}\frac{1}{r_0}}.\label{eq:ChemicalPotG4propX}
\end{equation}
The integration constant $M$ is related to $f_0$ in~\eqref{eq:BetterFormf} via $M=pf_0$, and it is also related to the physical regulated mass per unit area in the $(x^1,x^2)$ plane. Additionally, we have also defined $\hat{\gamma}\equiv\gamma p^2$, and ${D}$ denotes the Dawson function~\cite{abramowitz+stegun}:
\begin{equation}
    {D}(z)=\frac{1}{2}e^{-z^2}\sqrt{\pi}\,\mathrm{erfi}(z),
\end{equation}
with 
\begin{equation}
    \mathrm{erfi}\,z=-i \mathrm{erf}(iz),\qquad \mathrm{erf}\,z=\frac{2}{\sqrt{\pi}}\int_0^z\dd u\exp(-u^2).
\end{equation}
We take $G_4>0$ which amounts to $\gamma<0$. This is consistent with the reality of~\eqref{eq:FullsolutionG4propX}, and it further guarantees a non-negative Wald entropy.

Expanding at large $r$, we have 
\begin{eqnarray}
    h&\underset{r\to \infty}{=}&\frac{p}{r}k_{r\to\infty}-\frac{\Lambda r^2}{3}-\frac{2\hat{\gamma}\Lambda}{3}-\frac{M}{ r}+\mathcal{O}(r^{-2}),\label{eq:AsymptoticSeriesh}\\
    f&\underset{r\to \infty}{=}&\frac{p}{r}k_{r\to\infty}-\frac{\Lambda r^2}{3}-\frac{\hat{\gamma}\Lambda}{3}-\frac{M}{ r}+\mathcal{O}(r^{-2})\label{eq:AsymptoticSeriesf},
\end{eqnarray}
where $k_{r\to \infty}$ denotes the full asymptotic series expansion of $k$ with leading-order term $k_{r\to \infty}^{(0)}$. Note that $r\to \infty$ amounts to $\Bar{X}\to 0$. Let us assume that $k$ admits a power series expansion about the boundary with $k_{r\to \infty}^{(0)}\propto r^\Delta$. To leading power, the first terms in~\eqref{eq:AsymptoticSeriesh} and~\eqref{eq:AsymptoticSeriesf} grow as $\sim r^{\Delta-1}$. If $\Delta<3$ the asymptotic geometry is locally AdS once we fix $\Lambda=-3$, with metric:\footnote{Throughout this text we set $L_{UV}=1$.}
\begin{equation}
    \dd s^2_{UV}\sim\frac{1}{u^2}(-\dd t^2 +\dd r^2 +\delta_{ij}\dd x^i\dd x^j),\label{eq:AsymptoticFormG4propX}
\end{equation}
in terms of the holographic radial coordinate $u(r)\equiv r^{-1}$. If $\Delta=3$, then we observe that
\begin{equation}
    h,f\underset{r\to \infty}{\sim}-\frac{\Lambda_{eff}r^2}{3},
\end{equation}
and one can still achieve the asymptotic form~\eqref{eq:AsymptoticFormG4propX} provided $\Lambda_{eff}<0$.
On the other hand, if $\Delta>3$, we have that $g^{rr}\sim r^{\Delta-1}$ close to the boundary. In~\cite{Copsey:2012gw}, it was shown that this asymptotic behavior is tied to diverging curvature invariants as $r\to \infty$. This can be seen by writing the Riemann tensor in the static orthonormal frame, the diagonal tetrad, and asymptotically expanding it, only to find out that positive powers of the radial coordinate appear in its components. Consequently, any invariant built from the Riemann tensor will unavoidably diverge as $r\to \infty$, rendering the solution pathological. Especially from a holographic perspective, this is ultimately unwanted since gravitation would not become decoupled at the boundary. Therefore, one should only consider solutions with $\Delta\leq 3$.

So far, Eq.~\eqref{eq:FullsolutionG4propX} is a solution in models which may or may not contain kinetic terms for the axions, collectively denoted by $\Bar{X}$. To find the contribution of kinetic terms to the solution, we set $K=\bar{X}+\Tilde{K}$, and perform phase 2 in order to obtain:
\begin{equation}
    k=\tilde{k}-\frac{p}{2}\exp(\frac{\hat{\gamma}}{2 r^2})\pqty{r-\sqrt{2\Hat{\gamma}}{D}\pqty{\frac{\sqrt{\hat{\gamma}}}{r\sqrt{2}}}},
\end{equation}
with
\begin{equation}
    \tilde k\equiv \int^{\Bar{X}}\dd u\frac{\sqrt{C(u)}\tilde K(u)p^2}{4(-u)^{5/2}},\label{eq:tilkEquation}
\end{equation}
At large $r$,
\begin{equation}
    k\underset{r\to\infty}{=}\tilde{k}_{r\to\infty} -\frac{pr}{2}+\mathcal{O}(r^{-1}),\label{eq:AsymptoticExpansionKinetic}
\end{equation}
and thus, the presence of kinetic terms in~\eqref{eq:HorndeskiAction} corresponds to a contribution at the level of~\eqref{eq:FullsolutionG4propX}, which asymptotically is in $\mathcal{O}(1)$. We mention here that the highly relevant and well-studied solution in~\cite{Jiang:2017imk} follows from~\eqref{eq:FullsolutionG4propX} by choosing $K\propto\bar{X}$, or equivalently, $\tilde{K}=0$. Hence, we have provided a straightforward extension of the solutions in \cite{Jiang:2017imk} for literally any reasonable function $\tilde{K}$ with contribution $\tilde{k}$ given by~\eqref{eq:tilkEquation}.

There are many choices that yield a closed-form expression for the integral in the rhs of~\eqref{eq:tilkEquation}, and as an example, we will choose $\tilde{K}=\beta(-\bar{X})^n$ which leads to 
\begin{equation}
    \tilde{k}=\beta 2^{n-\frac{7}{2}}(-\gamma)^{\frac{3}{2}-n}p^2\Gamma\pqty{n-\frac{3}{2},-\frac{\hat{\gamma}}{2 r^2}}.
\end{equation}
Here, $\Gamma$ is the incomplete Gamma function~\cite{abramowitz+stegun} with the series expansion 
\begin{equation}
    \Gamma(\alpha,z)\underset{z\to 0}{=}\Gamma(\alpha)-\frac{z^\alpha}{\alpha}+\mathcal{O}(z^{\alpha+1}).
\end{equation}
Remember that we previously demanded $\Delta<3$. This amounts to $n>0$, and we have three asymptotic branches for $\tilde{k}$, namely, 
\begin{equation}
    \tilde{k}_{r\to \infty}^{(0)}\propto\beta
    \begin{cases}
    \Gamma\pqty{n-\frac{3}{2}},& n>\frac{3}{2}\\
    \log r,& n=\frac{3}{2}\\
    r^{3-2n},& 0<n<\frac{3}{2}
    \end{cases},
\end{equation}
where the proportionality factor is positive. Since $n>0$,~\eqref{eq:FullsolutionG4propX} can only describe a family of AAdS black branes when $\Lambda<0$, the horizon structure of which we are going to investigate in the remainder of this subsection.

\begin{figure}
     \centering
     \includegraphics{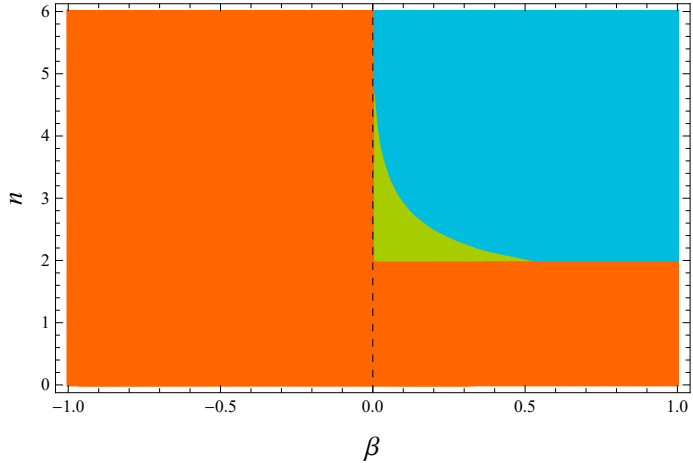}
     \caption{The space of AAdS black brane solutions on the $(n,\beta)$ plane at fixed $q_e=1=p$. Regions contain configurations where~\eqref{eq:ExtremalPolyG4propX} has one positive real root (orange), two positive real roots (green), no positive real roots (blue). The green region does not extend to $\beta=0$.}
     \label{fig:AAdSpop}
 \end{figure}

 Let us set $\Lambda=-3$ and obtain the Hawking temperature by evaluating~\eqref{eq:HawkingTemperature}:
 \begin{equation}
     T_H=-\exp(\frac{\hat{\gamma}}{2 r_0^2})\frac{q_e^2+2p^2r_0^2 -12 r_0^4 -2\beta p^{2n} r_0^{2(2-n)}}{16\pi r_0^3}
 \end{equation}
 The first step will be to determine the location of the extremal horizon by finding the largest solution to 
 \begin{equation}
      12 r^4 +2\beta p^{2n} r^{2(2-n)}-2p^2r^2-q_e^2=0\label{eq:ExtremalPolyG4propX}
 \end{equation}
 The above equation is just~\eqref{eq:ExtremalPoly} for $K=\bar{X}+\beta (-\bar{X})^n$, while it is also the equation for the extrema of an auxiliary function 
 \begin{equation}
     \mathcal{M}(r)\equiv rf(r)\sqrt{C(r)}+M.\label{eq:AuxFunction}
 \end{equation}
 This function will come in handy in the analysis that follows. We study~\eqref{eq:ExtremalPolyG4propX} numerically, finding cases  with even two positive real roots (see figure~\ref{fig:AAdSpop}). 
In general for $l$ distinct positive real roots, the largest of them will be the location of the event horizon of the extremal black brane, whereas the remaining $(l-1)$-many ones are expected to give rise to states with up to $l$-many inner horizons. Let us explicitly show this for the case at hand by using~\eqref{eq:AuxFunction}.

Assume there are two roots, $r_1$ and $r_2$, with $r_2>r_1$. Both are roots of $T_H$, and thus, are also double roots of $h,f$. Clearly, $r_e=r_2$. Additionally, they correspond to extrema of $\mathcal{M}$. The latter goes to $\infty$ as $r\to \infty$ by definition, while it goes to a finite value $\tilde{\mathcal M}$ as $r\to 0$, in particular,
\begin{equation}
    \tilde{\mathcal M}\equiv \frac{1}{4}\sqrt{-\frac{\pi}{2\hat{\gamma}}}\pqty{q_e^2+2\hat{\gamma}p^2-4\hat{\gamma}^2}.
\end{equation}
The extremal black brane has $M_e\equiv \mathcal{M}(r_e)$, with a naked singularity for all $M<M_e$. However, $f$ goes as 
 \begin{equation}
     f\underset{r\to 0}{\sim} \exp(-\frac{\hat{\gamma}}{2 r^2})\frac{\tilde{\mathcal{M}}-M}{r}+\frac{q_e^2}{4\hat{\gamma}}+...,
 \end{equation}
and if $\tilde{\mathcal{M}}<M_e$, it would develop a horizon for $\tilde{\mathcal{M}}<M<M_e$ which contradicts the previous statement. Therefore, $\tilde{\mathcal{M}}>M_e$, and moreover, $M_1>M_e$ where $M_1\equiv \mathcal{M}(r_1)$. Since $\mathcal{M}$ has two extrema for $r>0$, and $M_e$ is a global minimum, it has to be that $M_1$ is a local maximum, and thus, $M_e<\tilde{\mathcal{M}}<M_1$. This information is enough to conclude that: 
\begin{enumerate}[(i)]
    \item For $M<M_e$, the solution describes a naked singularity.
    \item For $M=M_e$, the solution describes an extremal black brane with its event horizon located at $r=r_e$.
    \item For $M_e<M\leq \tilde{\mathcal{M}}$, the solution describes a black brane which further develops an inner horizon. The causal structure is similar to that of Reissner-Nordstr\"om black holes. 
    \item For $\tilde{\mathcal{M}}<M<M_1$, the solution describes a black brane with two inner horizons. This is the picture of a ``black hole inside a black hole''~\cite{Martinez:2005di}.
    \item For $M=M_1$, the two inner horizons coalesce into one at $r=r_1$.
    \item For $M>M_1$, the solution describes a black brane with a regular event horizon.
\end{enumerate}
In the present work, we have not studied in detail the nature of the geometry inside the black hole horizon, but it would be interesting to perform such analysis on the lines of  \cite{Hartnoll:2020fhc,Cai:2020wrp,Grandi:2021ajl,Yang:2021civ}. It is reasonable to expect that the Horndeski terms play a role in such a game as already emphasized in \cite{Devecioglu:2021xug}.

All of these extremal black branes are finite-entropy domain wall solutions interpolating between a unit-radius $AdS_4$ in the UV and an $AdS_2\times \mathbb{R}^2$ IR geometry, where the size of $AdS_2$ can be numerically obtained for all cases via~\eqref{eq:AdS2size}. An exact example of this may be useful. Let us set $n=2$. For 
\begin{equation}
    12 q_e^2+(1-24\beta)p^4\geq 0,
\end{equation}
the location of the event horizon of the extremal $n=2$ AAdS black brane is located at 
\begin{equation}
    r_e=\bqty{\frac{1}{12}\pqty{p^2+\sqrt{12 q_e^2+(1-24\beta)p^4}}}^{1/2}.
\end{equation}
Then, using~\eqref{eq:AdS2size}, we find that 
\begin{equation}
    L_{IR}^2=\frac{1}{6}\pqty{1+\frac{p^2}{\sqrt{12 q_e^2+(1-24\beta)p^4}}}.
\end{equation}
The fact that there is no dependence on $\gamma$ is consistent with~\eqref{eq:zeroTempiiConstraint}, where the denominator simply becomes $r_e^4$. As a concluding remark, we mention here that we have performed a surfacial analysis of a very specific family in the $G_4\propto \bar{X}$ class, given by $K=\bar{X}+\beta(-\bar{X})^n$. There are many other choices $K=\bar{X}+\tilde{K}$ which still yield exact solutions, and it would be interesting to investigate them.

\subsection{\boldmath The $G_4\propto (-\bar{X})^m$ class}

As our second example, we set $G_4=\gamma(-\bar{X})^m$. Completing phases 1 and 3 in figure~\ref{fig:FlowchartFig}, we immediately arrive at the full bulk solution:
\begin{eqnarray}
    f&=&C^{\frac{3-4m}{2(2m-1)}}\bqty{-\frac{M-pk}{r}-\frac{\Lambda r^2}{3}{F}_{2,1}\pqty{{v,w\atop w+1};c_m} +\frac{q_e^2}{4 r^2}{F}_{2,1}\pqty{{v,-\frac{w}{3}\atop 1-\frac{w}{3}};c_m} },\nonumber\\
    C&=&(1-c_m)^{-2v},\qquad a=c_0-\frac{q_e}{r}{F}_{2,1}\pqty{{v,-\frac{w}{3}\atop 1-\frac{w}{3}};c_m},\qquad \phi^I=p\delta^I_i x^i.\label{eq:FullSolutionG4propMON}
\end{eqnarray}
Here $c_0$ is an integration constant related to the chemical potential $\mu$ in the dual field theory, while 
\begin{equation}
    c_m\equiv \gamma (m-1)\pqty{\frac{p}{r}}^{2m},\qquad v\equiv-1+\frac{1}{2(1-m)},\qquad w\equiv -\frac{3}{2m}.\label{eq:CustomNotation}
\end{equation}
The integration constant $M$ is related to $f_0$ in~\eqref{eq:BetterFormf} via $M=p f_0$, and again, it is also related to the mass (per unit area) of the solution. The symbol ${F}_{2,1}$ denotes Euler's hypergeometric function~\cite{abramowitz+stegun}: 
\begin{equation}
    {F}_{2,1}\pqty{ {\alpha,\alpha' \atop \alpha'' };z }\equiv{}_2F_1\pqty{\alpha,\alpha',\alpha'';z},\qquad \alpha,\alpha',\alpha''\in \mathbb{R},
\end{equation}
with 
\begin{equation}
    F_{2,1}\pqty{ {\alpha,\alpha' \atop \alpha'' };z }\underset{z\to 0}{=}1+\frac{\alpha\alpha' z}{\alpha''}+\mathcal{O}(z^2).\label{eq:Expansionz0}
\end{equation}
On the other hand, if $z\to \infty$, one needs to perform Pfaff's transformation~\cite{abramowitz+stegun} in order to send $z\to z^{-1}$, provided that the phase of $-z$ is less than $\pi$. Then, 
\begin{equation}
    F_{2,1}\pqty{ {\alpha,\alpha' \atop \alpha'' };z }\underset{z\to \infty}{=}(-z)^{-\alpha}\bqty{\frac{\Gamma(\alpha' -\alpha)\Gamma(\alpha'')}{\Gamma(\alpha''-\alpha)\Gamma(\alpha')}+\mathcal{O}(z^{-1})}+\alpha \leftrightarrow \alpha'.\label{eq:ExpansionzInf}
\end{equation}
The solution is well-defined for $C>0$ (which amounts to $c_m<0$) and all $m$ such that the hypergeometric functions are well-defined. In particular, the former restriction takes the form $m<1$ under the entropy requirement $G_4\geq 0$ (meaning $\gamma$>0), whereas the latter restriction can be shown to be
\begin{equation}
    m<-\frac{1}{2}\vee \pqty{m<0 \wedge \frac{1}{2m}\notin \mathbb{Z}}\vee \bqty{-\frac{3}{2m}\notin \mathbb{Z}\wedge m<1\wedge\pqty{m>0 \vee \frac{1}{2m}\notin \mathbb{Z}}}.\label{eq:ConditionI}
\end{equation}
Here, we used $\wedge,\vee$ as standard symbols for logical operations ``and'',``or'', respectively. This condition looks absolutely horrible and cryptic, but bear with us since will eventually visualize it in just a bit. 

\begin{figure}
    \centering
    \includegraphics{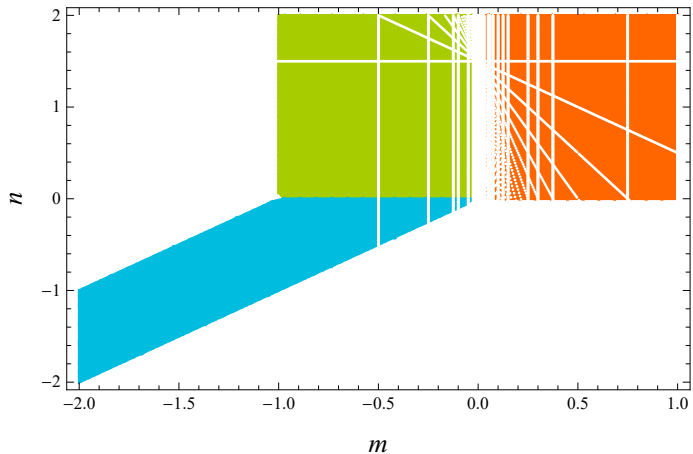}
    \caption{Three distinct classes of asymptotic forms as regions on the infinite $(m,n)$ plane, corresponding to leading-mode growth: $g^{rr},-g_{tt}\sim -\Lambda r^2$ (orange), $g^{rr}\sim -\Lambda r^{2(m+1)}$ and $-g_{tt}\sim -\Lambda r^{-2(m-\frac{1}{1-m})}$ (green), $g^{rr}\sim \beta r^{2(1+m-n)}$ and $-g_{tt}\sim \beta r^{\frac{2(1-n)-2m(1-m-n)}{1-m}}$ (blue). Regions extend indefinitely. White space represents the excluded region, after all previously discussed restrictions have been imposed.}
    \label{fig:MNplaneAllowed}
\end{figure}

Now, $z=c_m$ in the above expansions, and $c_m\to 0$ as $r\to \infty$ for $m>0$. Therefore, to leading order at large $r$, 
\begin{equation}
    h,f\underset{r\to \infty}{\sim} \frac{p}{r}k_{r\to \infty}-\frac{\Lambda r^2}{3}+...,
\end{equation}
and we may use the arguments of the previous subsection regarding the Riemann tensor to constrain the leading mode in $k_{r\to \infty}$. Practically, this means that for any positive $0<m<1$, we only consider AAdS solutions. Now, for $m<0$, $c_m\to \infty$ as $r\to \infty$, and the picture is quite different: 
\begin{eqnarray}
h&\underset{r\to \infty}{\sim}&-\frac{\Lambda (1-m)^{2-\frac{1}{1-m}}p^{-\frac{2m^2}{1-m}}\gamma^{-\frac{m}{1-m}}r^{-2\pqty{m-\frac{1}{1-m}}}}{3-2m(2-m)}+\frac{[\gamma(1-m)]^{-\frac{1}{2(1-m)}} r^{-2+\frac{1}{1-m}}}{p^{-2+\frac{1}{1-m}}}k_{r\to \infty},\nonumber\\
f&\underset{r\to \infty}{\sim}& -\frac{\Lambda p^{-2m} r^{2(1+m)}}{\gamma[3-2m(2-m)]}+\frac{[\gamma(1-m)]^{-2+\frac{1}{2(1-m)}}r^{4m-\frac{1}{1-m}}}{p^{4m-\frac{1}{1-m}}}k_{r\to \infty}.
\end{eqnarray}
Let us again assume that $k$ admits a power series expansion about the boundary with $k_{r\to \infty}^{(0)}\propto r^\Delta$. Here, the asymptotic form of the metric cannot have the AdS scaling, and thus, the only asymptotic geometry of interest (with constant curvature invariants at radial infinity) will be Lifshitz. However, before we start fixing parameters, let us completely determine the solution by choosing $K=\bar{X}+\beta(-\bar{X})^n$ and completing the missing phase 2 in figure~\ref{fig:FlowchartFig}. Doing so, we find the contribution:
\begin{equation}
    k=-\frac{pr}{2}F_{2,1}\pqty{{v,\frac{w}{3}\atop 1+\frac{w}{3}};c_m}+\frac{\beta p^{2n-1}r^{3-2n}}{2(3-2n)}F_{2,1}\pqty{{v,w\pqty{1-\frac{2n}{3}}\atop 1+w\pqty{1-\frac{2n}{3}}};c_m },\label{eq:kcontrib}
\end{equation}
with $c_m,v,w$ defined in~\eqref{eq:CustomNotation}. This particular form implies the identification: 
\begin{equation}
    \Delta = \begin{cases}
    1,& 1>m>0,n\geq 1\\
    3-2n,& 1>m>0,n<1\\
   1-2m\pqty{1-\frac{1}{2(1-m)}},& m<0,n\geq 1 \\
    3-2n-m\pqty{2-\frac{1}{1-m}},& m<0,n<1
    \end{cases}.
\end{equation}
For~\eqref{eq:kcontrib} to be well-defined, we need to impose further restrictions which we do not write down explicitly; basically, we take the union of these conditions with~\eqref{eq:ConditionI}, and we depict the logical negation of that union as white space in figure~\ref{fig:MNplaneAllowed}, together with exclusions coming from filtering out the cases with diverging tidal forces close to the boundary. Most importantly, in figure~\ref{fig:MNplaneAllowed} we display the three distinct kinds of asymptotic forms we end up with. Notice that we have severly restricted the $(m,n)$ parameter space.

As previously anticipated, for negative $m$ there exists a family of asymptotic Lifshitz (ALif) geometries. We will show that this corresponds to the lower boundary of the blue region in figure~\ref{fig:MNplaneAllowed}. Lifshitz spacetimes are described by the metric~\cite{Kachru:2008yh}:
\begin{equation}
    \dd s^2=-{r^{2z}}\dd t^2+\frac{\dd r^2}{r^2}+r^2\delta_{ij}\dd x^i\dd x^j,\label{eq:LifshitzMetric}
\end{equation}
where $z$ is the dynamical critical exponent, and we have set $L=1$. The key-feature of~\eqref{eq:LifshitzMetric} is its symmetry under the anisotropic scale transformations
\begin{equation}
    t\to \lambda^z t,\qquad r\to \lambda^{-1}r,\qquad x^i\to \lambda x^i.\label{eq:LifTF}
\end{equation}
Here, we expect our black branes to be locally asymptotically of the form~\eqref{eq:LifshitzMetric} with $z$ being a function of $m,n$, under certain conditions. First, if $h$ is going to grow faster than $f$, it has to be $m<0$. Then, if $f$ is to grow as $\sim r^2$, it further has to be that $n<0$ and $m=n$. As we have set $L=1$, the coefficient of the $g^{rr}$ leading mode has to be set to unity which implies 
\begin{equation}
    \beta=2\gamma (3-6m+4m^2).\label{eq:betafix}
\end{equation}
This ensures that $g^{rr}\sim r^2(1+...)$ at large $r$, where the dots denote terms in orders of growth less than $\mathcal{O}(1)$. Having secured the desired growth for $f$, we can identify
\begin{equation}
    z(m)\equiv \frac{1}{1-m}-2m,\label{eq:Identifyz}
\end{equation}
with $z>1$ amounting to $m<0$, which is exactly the branch we are studying. Additionally, using~\eqref{eq:betafix} and setting
\begin{equation}
    \gamma=\frac{p^{-2m}}{1-m},
\end{equation}
we can make sure that $-g_{tt}\sim r^{2z}(1+...)$ as $r\to \infty$. Summarizing, the tuning:
\begin{equation}
    \beta=\frac{2(3-6m+4m^2)p^{-2m}}{1-m},\qquad \gamma=\frac{p^{-2m}}{1-m},\label{eq:LifshitzTunings}
\end{equation}
together with the identification:
\begin{equation}
    m=\frac{1}{4}\pqty{2-z+\sqrt{4(z-1)+z^2}},\label{eq:mofz}
\end{equation}
which is~\eqref{eq:Identifyz} inverted, give rise to axionic black branes with a metric locally acquiring the asymptotic form~\eqref{eq:LifshitzMetric}. To the best of our knowledge, this is the only exact Lifshitz-like family in Horndeski gravity with arbitrary $z>1$,\footnote{Lifshitz-like black holes in Horndeski theory have been also reported in~\cite{2013arXiv1312.7736B} and~\cite{Brito:2019ose}, but in a different context. The authors of the former used a scalar $\phi(t,r)$, and constructed an exact ALif solution with a specific $z=1/3$, whereas in the latter work, the solution is for $z=1/2$ and $\phi\equiv \phi(r)$. Notice that both cases are for $z<1$.} where $z$ is fixed once the degree of the $G_4$ monomial is chosen.

\begin{figure}
    \centering
    \includegraphics{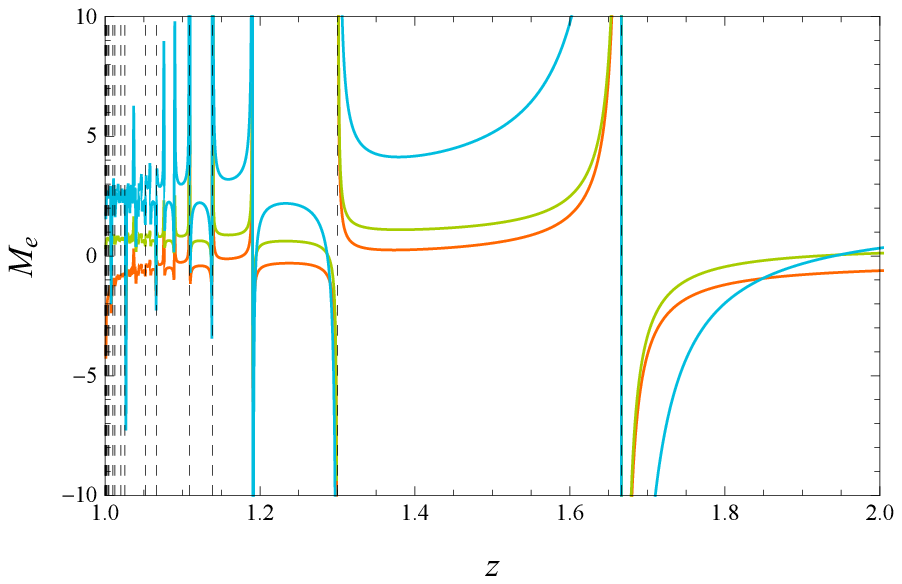}
    
    \vspace{0.4cm}
    
    \includegraphics{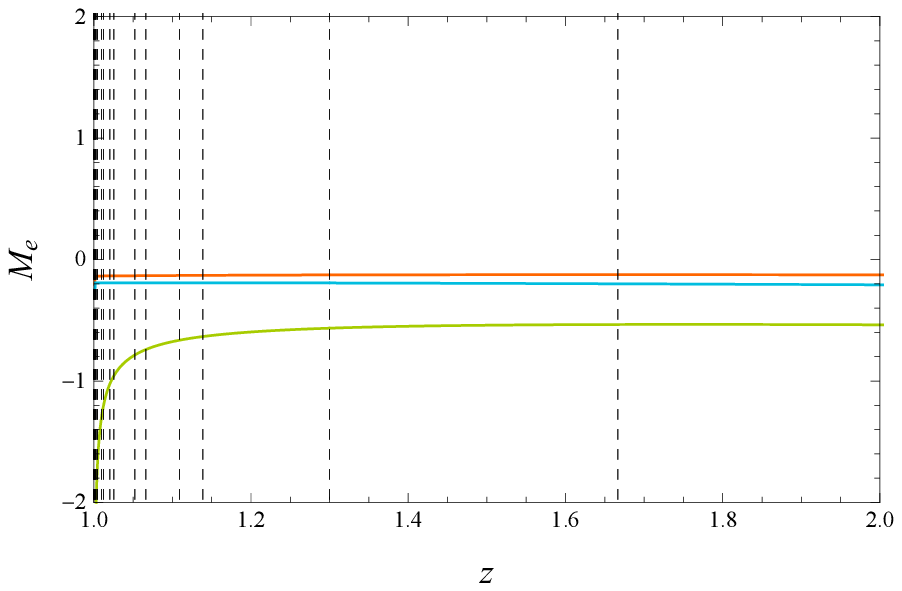}
    \caption{The extremal mass $M_e$ as a function of $z$. \textbf{Top panel:} $q_e=1,\Lambda=-3$ (green), $q_e=2,\Lambda=-3$ (blue), and $q_e=1,\Lambda=3$ (orange). \textbf{Bottom panel:} The neutral case $q_e=0$ and $\Lambda=-3$ (red), $\Lambda=0$ (blue), $\Lambda=3$ (green). Vertical dashed lines denote $z$ discontinuities corresponding to points outside the domain of the real-valued hypergeometric functions in the solution.}
    \label{fig:LifshitzMe}
\end{figure}

Now, we turn our attention to (deep) IR geometries which can act as ground states. One can show that for~\eqref{eq:FullSolutionG4propMON}, it is impossible to construct a regular core  near the origin of the radial coordinate; the curvature invariants exhibit poles which cannot be removed by tuning. Hence, the only candidate left is the extremal IR geometry~\eqref{eq:IRmetric}. As we discussed in the previous section, the location at which the extremal black branes develop a horizon, is the same for all $G_4$ classes with a common choice for $K$, a fact which saves us a lot of time, since we have already performed this analysis in the last subsection where we also chose $K=\bar{X}+\beta (-\bar{X})^n$. In the current case, it makes sense to extend figure~\ref{fig:AAdSpop} to the negative $n$ half plane which is occupied by configurations where~\eqref{eq:ExtremalPolyG4propX} has one positive real solution corresponding to the location of the extremal horizon. Apart from the usual $AdS_4\to AdS_2\times \mathbb{R}^2$ flow for $1>m>0,\,n>0$, here we also have extremal finite-entropy bulk geometries interpolating between a Lifshitz asymptotic geometry and $AdS_2\times \mathbb{R}^2$ in the infra-red. For these zero-temperature configurations, we numerically determine the extremal value of $M$, $M_e(z)$, plotting it in figure~\ref{fig:LifshitzMe}.
Close to $z=1$, the $z$ discontinuities become more frequent, and $M_e$ exhibits a peculiar oscillating behavior. As $z$ grows, $M_e$ goes to some constant value. These wild fluctuations also cease to exist once we go to the electrically neutral version of these extremal branes, as can be seen in figure~\ref{fig:LifshitzMe}.
Overall, the charge seems to have an uplifting role, whereas a positive $\Lambda$ seems to have the opposite effect, pushing $M_e$ to lower values. Note that $\Lambda\geq 0$ is allowed in this case, since the bulk cosmological constant is unrelated to the locally asymptotically Lifshitz form of the metric, a fact made obvious by its absence in the tuning~\eqref{eq:LifshitzTunings}. Also, negative values for $M_e$, and thus, negative values for $M>M_e$, is a known feature for planar solutions with axionic charges (see for example \cite{Bardoux:2012aw}).

\begin{figure}
    \centering
    \includegraphics{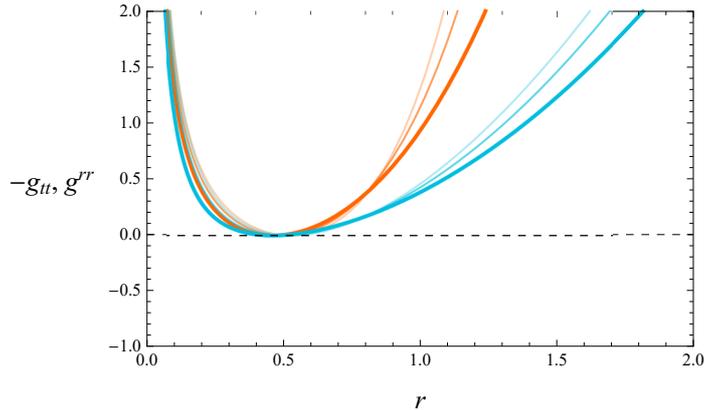}
    \caption{Plotting $-g_{tt}$ (orange) and $g^{rr}$ (blue) versus the radial coordinate $r$ for \emph{neutral} ALif extremal black branes with $\Lambda=0$ at $p=1$. From high to low opacity, $z=3/2$, $z=2$ and $z=5/2$, respectively.}
    \label{fig:ExtremalLifBranes}
\end{figure}

Let us close this subsection with the easiest exact example, the case with $q_e=0=\Lambda$. The extremal ALif black brane will develop a horizon at 
\begin{equation}
    r_e=\pqty{\frac{2+3z+\sqrt{4(z-1)+z^2}}{p^2}}^{-\frac{2}{2+z+\sqrt{4(z-1)+z^2}}},
\end{equation}
the latter supported by the axion flux $p$. This is the exact solution to~\eqref{eq:ExtremalPolyG4propX}. The near-horizon metric will have the form~\eqref{eq:IRmetric}, with the size of the IR $AdS_2$ given by 
\begin{equation}
    L_{IR}^2=\frac{1+(1-m)^{-1+\frac{1}{1-m}}(6-4m(3-2m))^{-\frac{m}{1-m}}p^{\frac{2m}{1-m}}}{3-6m+4m^2},\label{eq:LIRlifshitzcase}
\end{equation}
obtained by evaluating~\eqref{eq:AdS2size},\footnote{For the general case, the length can only be numerically determined using~\eqref{eq:AdS2size}.} where we expressed the result in terms of $m(z)$ (see~\eqref{eq:mofz}). The behavior of the two metric functions, $h$ and $f$, can be seen in figure~\ref{fig:ExtremalLifBranes}.
The full (extremal) solution is then interpolating between a UV geometry~\eqref{eq:LifshitzMetric} with the Lifshitz scaling~\eqref{eq:LifTF}, at least up to some appropriate cut-off, and an IR geometry~\eqref{eq:IRmetric} with the radius (of curvature) of $AdS_2$ given by the square root of~\eqref{eq:LIRlifshitzcase}, while its entropy is also finite:
\begin{equation}
   \tilde{s}=2\pi p^{\frac{2}{1-m}}\frac{(1-m)^{\frac{1}{1-m}}\bqty{6-4m(3-2m)}^{-\frac{m}{1-m}}+p^{-\frac{2m}{1-m}}}{3-6m+4m^2}.
\end{equation}
As a concluding statement, we also remark here that~\eqref{eq:FullSolutionG4propMON} matches the solution in~\cite{Figueroa:2020tya} for $m=1/2$ and $K=\bar{X}$, at least up to some insignificant arithmetic differences, which are due to the slightly different action principle used there.

\section{Holographic DC conductivities}
\label{sec5}
In this section, we consider a general black-brane background like those in the previous section with the further requirement that the UV fixed point of the dual field theory is relativistic invariant, i.e., the bulk geometry is (locally) asymptotically anti-de Sitter. We assume the presence of an event horizon at some radius $r=r_0$. We use an Eddington-Finkelstein (EF) coordinate system
\begin{equation}
    t=v-\frac{1}{4\pi T_H}\ln(\frac{r}{r_0}-1),\label{eq:EFcoords}
\end{equation}
in which the regularity of the metric~\eqref{eq:MetricAnsatz} in the vicinity of the horizon is manifest. The Hawking temperature is given by
\begin{equation}
    T_H=\frac{\sqrt{h'(r_0)f'(r_0)}}{4\pi},
\end{equation}
A finite chemical potential, defined as
\begin{equation}
    \mu=\int_{r_0}^{\infty} F_{rt}\, dr=a(\infty)\qquad (\text{with}\,a(r_0)=0),
\end{equation}
is introduced in the dual field theory. In the rest of this section, we will nevertheless work in the thermodynamic ensemble where the charge density is kept fixed, instead of the chemical potential. The electric charge density is given by
\begin{equation}
\ev{J^t}=\lim\limits_{r\to \infty}\sqrt{\frac{f}{h}}a' r^2=q_e,
\end{equation}
where in the last equation we used~\eqref{eq:ElectricField}. 

To calculate the conductivities we closely follow~\cite{Donos:2014cya}, and consider the following linearized perturbations about the class of backgrounds we previously presented:
\begin{align}
    \var g_{tx^1}(t,r)&=H_{tx^1}(r)+t g_{2}(r),\\
    \var g_{rx^1}(r)&=r^2H_{rx^1}(r),\\
    \var A_{x^1}(t,r)&=a_{x^1}(r)+t g_{1}(r),\\
    \var \phi^1(r)&=\chi(r).
\end{align}
We also once again write $h=Cf$. Looking at the Maxwell equations of motion, at linear level there is only one nontrivial component, the one in the $x^1$ direction:
\begin{equation}
    \partial_r\pqty{\sqrt{-g}F^{rx^1}}=0,
\end{equation}
where, using the background equations,
\begin{equation}
    \sqrt{-g}F^{x^1r}=-\frac{H_{tx}a'+ha_x'}{\sqrt{C}} - \frac{t(g_2 a' +h g_1')}{\sqrt{C}}\equiv J^{x^1} - \frac{t(g_2 a' +h g_1')}{\sqrt{C}}.\label{eq:x1CompGauge}
\end{equation}
Here, $J^{x^1}$ represents the time-independent electric current in the $x^1$ direction when evaluated at $r\to \infty$. Notice that time dependence will drop out from~\eqref{eq:x1CompGauge}, if 
\begin{equation}
    g_2=-\frac{hg_1'}{a'}.\label{eq:zerotime1}
\end{equation}
We will keep this in mind. Turning our attention to the linearized Einstein equations, we have two nontrivial components, $tx^1$ and $rx^1$. Considering $\mathcal{E}_{tx^1}$ first, and by using the (linearized) Maxwell equations of motion \eqref{eq:x1CompGauge}, we can show that 
\begin{equation}
    -2\sqrt{C}\mathcal{E}_{tx^1}=\gamma_{0}\var g_{tx^1}+\gamma_{1}\var g'_{tx^1}+\gamma_{2}\var g''_{tx^1}-(\sqrt{-g}a F^{x^1r})'=0,\label{eq:precon}
\end{equation}
where the $\gamma_{n}$'s are complicated functions of $r$. Their explicit form is irrelevant to our purpose for the moment. The key step here is to manifestly write this equation as $\partial_r\tilde{Q}=0$ which amounts to $\tilde{Q}$ being $r$-independent. 

The concrete expression for $\tilde{Q}$ turns out to be: 
\begin{equation}
    \tilde{Q}\equiv U(r)(V(r)\var g_{tx^1})'-\sqrt{-g}a F^{x^1r},\label{eq:guess}
\end{equation}
where $U$ and $V$ are to be determined by subtracting~\eqref{eq:precon} from the radial derivative of~\eqref{eq:guess} and solving the result order by order in derivatives of $\var g_{tx^1}$. Doing so, we first find that the vanishing of the coefficient in front of $\var g_{tx^1}''$ forces the relation
\begin{equation}
    U=\frac{\gamma_2}{V}.
\end{equation}
Using this information, we can deduce that the vanishing of the coefficient in front of $\var g_{tx^1}'$ can only be satisfied if 
\begin{equation}
    (\ln V)'+(\ln \gamma_2)'=\frac{\gamma_1}{\gamma_2}.
\end{equation}
Since the $\gamma_n$'s contain Horndeski functions, it will be convenient to write everything in terms of $\bar{X}$, and only then integrate, finding the solution
\begin{equation}
    V=\frac{1+G_4-\bar{X}G_{4\bar{X}}}{\gamma_2 \sqrt{C}},\qquad \gamma_2=\sqrt{hf}(1+G_4-\bar{X}G_{4\bar{X}}).
\end{equation}
This also guarantees that the coefficient in front of $\var g_{tx^1}$ vanishes, and results in
\begin{eqnarray}
    \tilde{Q}&=&\sqrt{-g}\frac{hf(1+G_4-\bar{X}G_{4\bar{X}})}{r^2}\pqty{\frac{\var g_{tx^1}}{h}}'-\sqrt{-g}a F^{x^1r}\nonumber\\
    &=&Q^{x^1}-aJ^{x^1}+\frac{t}{\sqrt{C}}\pqty{g_2 a' + h g_1' +(1+G_4-\bar{X}G_{4\bar{X}})(h g_2' -g_2 h')},
\end{eqnarray}
where
\begin{equation}
    Q^{x^1}\equiv \sqrt{-g}\frac{hf(1+G_4-\bar{X}G_{4\bar{X}})}{r^2}\pqty{\frac{ H_{tx^1}}{h}}',
\end{equation}
is the time-independent piece of the thermal current in the $x^1$ direction when evaluated at the boundary. Also taking into account~\eqref{eq:zerotime1}, the time dependence of $\tilde{Q}$ completely drops out if we choose 
\begin{equation}
    g_1=\zeta a + c_1,\qquad g_2=-\zeta h.\label{eq:notimeChoice}
\end{equation}
We set $c_1=-E$ with $E$ parametrizing an electric field deformation. On the other hand, $\zeta$ parametrizes a time-dependent heat source, a thermal gradient $\sim \nabla T$. Henceforth, $g_1,g_2$ as in~\eqref{eq:notimeChoice}.

We also need to solve the $rx^1$ component of the linearized Einstein equations. This is an expression algebraic in the $rx^1$ mode which has the solution:
\begin{equation}
    H_{rx^1}=\frac{\chi'}{p}+\frac{W}{fZp^2},\label{eq:hrxmode}
\end{equation}
where 
\begin{eqnarray}
    {W}&=&-Ea'+\zeta\bqty{aa'-h'(1+G_4-\bar{X}G_{4\bar{X}})+\frac{2h}{r}\pqty{1+G_4-2\bar{X}(G_{4\bar{X}}+\bar{X}G_{4\bar{X}\bar{X}})}},\quad\\
    Z&=&C K_{\bar{X}}-\frac{2G_{4\bar{X}}(h+rh')}{r^2}+\frac{2p^2G_{4\bar{X}\bar{X}}(2h+rh')}{r^4}.
\end{eqnarray}
Observe that at $r=r_0$, we have 
\begin{eqnarray}
    W&\underset{r=r_0}{=}&-E a'-{4\zeta \pi T_H\sqrt{C}(1+G_4-\bar{X}_0G_{4\bar{X}})},\\
    Z&\underset{r=r_0}{=}& C K_{\bar{X}}-\frac{8\pi T_H\sqrt{C}(G_{4\bar{X}}+\bar{X}_0G_{4\bar{X}\bar{X}})}{r_0},
\end{eqnarray}
all functions evaluated at $r_0$. Clearly, these are finite results, meaning that $H_{rx^1}$ will diverge as $\sim f^{-1}$ when $r\to r_0$, provided that $\chi$ is smooth analytic there. Note also that~\eqref{eq:hrxmode} solves the equation of motion for $\chi$. 

We must now ensure that all perturbations are well-behaved in the bulk, which means that certain boundary conditions have to be imposed, both at the horizon of the black brane, as well as at infinity. Starting with boundary conditions as $r\to \infty$, from the gauge equations of motion, we see that a sufficient fall-off for $a_{x^1}$ is $\sim J^{x^1}r^{-1}$ provided that $H_{tx^1}$ does not yield a non-normalizable mode. The last requirement ensures that only the time-dependent piece of $\var g_{tx^1}$ acts as a holographic source for the heat current at $r\to \infty$. Close to the conformal boundary, $\mathcal{E}_{tx^1}=0$ has two independent solutions, yielding 
\begin{equation}
    H_{tx^1}\underset{r\to \infty}{\sim} c_{+}w_+(\gamma,\Delta) r^2 + \frac{c_- w_{-}(\gamma,\Delta)}{r}F_{2,1}\pqty{{1,-\frac{3}{\Delta}\atop \frac{\Delta-3}{\Delta}};-\frac{\gamma(\Delta+2)r^\Delta}{2}},\label{eq:HtxBoundary}
\end{equation}
where $c_{\pm}$ are constants of integration, $w_{\pm}$ are some functions of $\gamma,\Delta$, and we assumed that $G_4$ can be expanded in power series about infinity with leading mode $\propto \gamma r^\Delta$, together with $f,h\sim r^2$, in order to get the above result. For $\Delta<0$, we can expand the hypergeometric function using~\eqref{eq:Expansionz0}; we get a non-normalizable mode $\propto c_+ r^2$ which we kill by setting $c_+=0$, and  a normalizable mode with fall-off in $\mathcal{O}(r^{-1})$. This behavior associates with a $G_4$ which goes to infinity as some positive power of $\bar{X}$. Also, for~\eqref{eq:HtxBoundary} to be well defined in this case, $\gamma(\Delta+2)\geq 0$. Assuming $\gamma$ is a coupling constant in front of the quartic sector, we saw that the non-negative entropy condition suggests that its sign is fixed positive.\footnote{For $G_4=\gamma X$, we said $\gamma<0$. This is consistent with the statements in this paragraph, since in that case $G_4$ would approach infinity as $\sim -\gamma r^{-2}$.} This means that $-2<\Delta<0$. But what about $\Delta>0$? Assuming $K$ also admits a power series expansion about the boundary, led by $\propto \beta r^\Theta$, we find that satisfying the bulk equations of motion at an $AdS_4$ UV fixed point, amounts to $\Delta,\Theta<0$. Therefore, we cannot have $\Delta>0$ and AdS asymptopia together, a fact which ensures that the fall-off of $H_{tx^1}$ is indeed $\sim r^{-1}$. It is satisfactory to verify (as a cross-check) that the asymptotic power restrictions we found here are consistent with those found for the AAdS black branes in section~\ref{sec:Examples}. No additional constraints are coming from the asymptotic behavior of $\var g_{rx^1}$:
\begin{equation}
    \var g_{rx^1}\underset{r\to \infty}{\sim} \frac{\chi'r^2}{p}+\frac{2q_e(-E+\zeta \mu)-2\gamma\zeta\Delta(\Delta+1)r^{\Delta+3}}{\beta\Theta r^{\Theta+4}+2\gamma\Delta(1+2\Delta) r^{\Delta+4}},
\end{equation}
It can fall off as fast as we wish, with the non-normalizable mode of $\chi$ vanishing and its normalizable piece in some arbitrary order of growth strictly less than $\mathcal{O}(r^{-1})$.

It remains to discuss the behavior of the perturbations at the horizon $r=r_0$. To do so we switch to EF coordinates $(v,r)$ with $v$ given by~\eqref{eq:EFcoords}. Considering the gauge-field perturbation first, 
we have that 
\begin{equation}
    \var A_{x^1}=a_{x^1}+(-E+\zeta a)\pqty{v-\frac{1}{4\pi T_H}\ln(\frac{r}{r_0}-1)}.
\end{equation}
Expanding about the horizon, we see that 
\begin{equation}
    a_{x^1}\underset{r\to r_0}{=} -\frac{E}{4\pi T_H}\ln\pqty{\frac{r}{r_0}-1}+\mathcal{O}(r-r_0),
\end{equation}
for $\var A_{x^1}$ to be regular there. A valid near-horizon approximation is then 
\begin{equation}
    a'_{x^1}\sim -\frac{E}{\sqrt{C}f}.\label{eq:axNEARapprox}
\end{equation}
Moreover, for $\delta g_{tx^1}$ to be regular at the black brane horizon, it must be that 
\begin{equation}
    H_{tx^1}\underset{r\to r_0}{\sim} U - \frac{\zeta h}{4\pi T_H}\ln(\frac{r}{r_0}-1)+...,
\end{equation}
where $U$ is some analytic function at $r=r_0$ with
\begin{equation}
    U\underset{r=r_0}{=}\sqrt{hf}r^2H_{rx}\mid_{r=r_0},
\end{equation}
assuming $\chi$ is constant there. Note that in order to find this result, we used the approximation~\eqref{eq:axNEARapprox} and the background equations of motion. In this way, the divergence in~\eqref{eq:hrxmode} is also remedied, and we finally have a set of well-behaved perturbations in the whole bulk. 

We can proceed with evaluating $J^{x^1}$ and $Q^{x^1}$ at the horizon radius, denoting the values as $\ev{J}$ and $\ev{Q}$, respectively: 
\begin{eqnarray}
    \ev{J}&=&E\pqty{1+\frac{C q_e^2}{Zr_0^2 p^2}}+\zeta \frac{4\pi q_e C T_H (1+G_4-\bar{X}_0G_{4\bar{X}})}{Z p^2},\\
    \ev{Q}&=&E \frac{4\pi q_e C T_H (1+G_4-\bar{X}_0G_{4\bar{X}})}{Z p^2}+\zeta  \frac{16\pi^2 r_0^2 C T_H^2 (1+G_4-\bar{X}_0 G_{4\bar{X}})^2}{Z p^2},
\end{eqnarray}
where all appearing functions are evaluated at $r=r_0$. From the above expressions we can read off the components of the holographic DC transport matrix:
\begin{equation}
    \pmqty{\sigma&\alpha T\\ \bar{\alpha} T&\bar{\kappa} T}\equiv \pmqty{\partial_E \ev{J}& \partial_\zeta \ev{J}\\ \partial_E \ev{Q}&\partial_\zeta \ev{Q}},
\end{equation}
where $\partial_E\equiv \pdv*{}{E}$ and $\partial_\zeta \equiv \pdv*{}{\zeta}$. At temperature $T=T_H$, the electric DC conductivity $\sigma$, the thermoelectric conductivities $\alpha,\bar{\alpha}$ and the heat conductivity (at zero electric field) $\bar{\kappa}$, are given by\footnote{Henceforth, we omit the $H$ subscript from $T_H$ for convenience. It should be understood that $T$ is always identified with the Hawking temperature of the black brane.} 
\begin{eqnarray}
    \sigma&=& 1+\frac{C q_e^2}{Zr_0^2 p^2},\label{eq:eleDC}\\
    \alpha&=& \frac{4\pi q_e C  (1+G_4-\bar{X}_0G_{4\bar{X}})}{Z p^2}=\bar{\alpha},\label{eq:ThermoEleDC}\\
    \bar{\kappa}&=&   \frac{16\pi^2 r_0^2 C T (1+G_4-\bar{X}_0 G_{4\bar{X}})^2}{Z p^2}\label{eq:HeatZeroE},
\end{eqnarray}
respectively. 

Notice that at zero temperature we have 
\begin{equation}
    \sigma \underset{T=0}{=}1+\frac{q_e^2}{K_{\bar{X}}(\bar{X}_e)r_e^2 p^2},
\end{equation}
which implies the presence of a resistivity residue. Note that the denominator of the coherent part of $\sigma$ is related to the (effective) graviton mass evaluated at $r=r_0$ \cite{Vegh:2013sk,Davison:2013jba,Baggioli:2014roa}, thus it better be that $Z(r_0)>0$, from which it immediately follows that $K_{\bar{X}}>0$ at zero temperature. Moreover, for vanishing heat flows $\ev{Q}=0$, only the incoherent part of $\sigma$ survives, and thus the conductivity is constant, equal to unity. This is made obvious by the fact that
\begin{equation}
    \ev{J}\underset{\ev{Q}=0}{=}E\pqty{\sigma-\frac{\alpha\bar{\alpha}T}{\bar{\kappa}}}=E,
\end{equation}
Additionally, the heat conductivity at zero electric current, $\kappa$, is given by 
\begin{equation}
    \kappa=\frac{16\pi^2 r_0^4 C T(1+G_4-\bar{X}_0G_{4\bar{X}})^2}{C q_e^2 + Z p^2 r_0^2};
\end{equation}
it can be found by acting with a $\zeta$ derivative on 
\begin{equation}
    \frac{\ev{Q}}{T}\underset{\ev{J}=0}{=} \zeta (\bar{\kappa} - \frac{\alpha\bar{\alpha}T}{\sigma}).
\end{equation}

Importantly, the discussion just presented is valid only when the scalar operators dual to the axionic fields $\phi^I$ break the translational invariance of the dual field theory explicitly, causing momentum dissipation. This is in 1-to-1 correspondence with saying that the bulk solution $\phi^I \sim x^I$ plays the role of an external source for the dual operator. As already explored in the literature \cite{Alberte:2017oqx}, this is not necessarily the case. In the opposite scenario, where translations are spontaneously broken, the corresponding DC conductivities would be clearly infinite and must be treated in a separate way not analyzed here.

As an example, let us focus on a specific family of models given by $G_4=\gamma(-\bar{X})^m$ and $K=\bar{X}+\beta (-\bar{X})^n$. Remember that for these choices, we found AAdS black branes in section~\ref{sec:Examples}, provided that $0<m<1$ and $n,\gamma>0$. We are particularly interested in configurations supporting an extremal profile, meaning that we should avoid the blue region in figure~\ref{fig:AAdSpop}.\footnote{We repeat here that this figure holds true for all $G_4$ classes with $K=\bar{X}+\beta (-\bar{X})^n$.} Here, the crucial function $Z$ takes the form:
\begin{equation}
    Z\underset{r=r_0}{=}C\bqty{1-n \beta \pqty{\frac{p}{r_0}}^{2(n-1)}}+{8\pi\gamma m^2 T\sqrt{C}}p^{2(m-1)}r_0^{1-2m}.\label{eq:Zspecific}
\end{equation}
If $\beta<0$, it is guaranteed that $Z$ is positive. For $\beta>0$, one first needs to wisely choose $n$ (based on the value of $\beta$) so that the extremal black brane exists. Then, a further condition needs to be imposed, namely, 
\begin{equation}
    \beta<\frac{r_e^{2(n-1)}}{np^{2(n-1)}},
\end{equation}
so that at $T=0$ we have $Z(r_e)>0$. However, eq.~\eqref{eq:Zspecific} should be positive for all $T>0$, equivalently all $r_0>r_e$, and therefore, this can only be true if $n>1$ regardless of $m,\gamma$, or if further conditions are imposed, relating $n,\beta$ with $m,\gamma$.  

The very large temperature regime is unfortunately not interesting since our model is conformal therein, and it cannot take into account the granularity of matter, for example, the presence of a well-defined lattice spacing. The electric conductivity displays a lower bound, i.e., $\sigma\geq 1$, which holds for all theories with bulk electrodynamics dictated by a pure Maxwell term \cite{Grozdanov:2015qia} but that can be easily violated otherwise \cite{Baggioli:2016oqk,Baggioli:2016oju}. Notice that at large temperatures ($T$ being the largest scale in the system compared to $q_e,p,\beta,\gamma$) the DC conductivity is universally $\sigma=1$. This is just a trivial outcome of the fact that the conformal UV fixed point is reached. It indeed simply coincides with the value one would obtain in a Schwarzschild-AdS background. On the other hand, for small and/or intermediate temperatures, we find a more interesting picture. First of all, the temperature is a strictly increasing function of $r_0$, that is to say, it is a positive bijective function for all $r_0>r_e$ which can in theory be inverted to give $r_0(T)$. However, for arbitrary $m,n$ we are unable to provide a closed-form expression of the latter, so we will work with parametric plots. 

\begin{figure}
    \centering
    \includegraphics{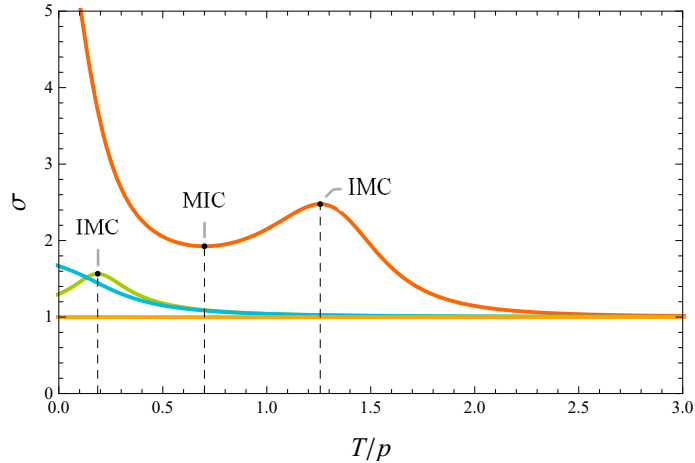}
    \caption{Electric DC conductivity $\sigma$ as a function of the dimensionless temperature $T/p$ for $q_e/p^2=1$, $\gamma=0.14$, $m=0.13$ and: $\beta=0.78$, $n=0.8$ (orange - class (iii)), $\beta=-0.2$, $n=4$ (green - class (ii)), $\beta=-0.5$, $n=2$ (blue - class (i)). The yellow line indicates the DC conductivity in the absence of heat flows which is also universally bounding $\sigma$ from below. Although not obvious in the plot, the orange curve does also meet the $\sigma$ axis at $T=0$ with $\sigma(0)\approx 7.37$.}
    \label{fig:DCconductivity}
\end{figure}

A numerical investigation (see figure~\ref{fig:DCconductivity}) reveals that the DC conductivity can be classified into three distinct categories based on the behavior at low/intermediate $T/p$. These are characterized by: (i) a persistent metallic phase in which $d \sigma/dT <0$ for any $T$, (ii) an insulator-metal crossover (IMC) where at small temperatures the DC conductivity exhibits a maximum (similarly to what some of us found originally in \cite{Baggioli:2014roa}) and (iii) a metal-insulator crossover (MIC) at low $T$, followed by an IMC at some intermediate larger scale $T/p\sim 1$. 

We stress out that one should take the word ``insulator'' with a grain of salt here. Using the strict condensed matter definition, an insulator is a system for which at $T=0$ the electric conductivity is zero. This is obviously not the case in this system because of the presence of the residual incoherent conductivity discussed above. Here, we operationally define an insulating state as a phase in which $d \sigma/dT>0$. To find the transition points (IMC and MIC), we write~\eqref{eq:eleDC} purely in terms of $r_0$, and differentiate the former with respect to the latter. Since the temperature is strictly increasing, we may use the chain rule
\begin{equation}
    \pdv{\sigma}{T}=\pdv{\sigma}{r_0}\pqty{\pdv{T}{r_0}}^{-1},
\end{equation}
to establish a 1-1 relation between each critical point, $T_k$, of $\sigma(T)$ and some unique positive real root of $\pdv*{\sigma}{r_0}$, $r_0(T_k)>r_e$. The three classes are represented in figure~\ref{fig:DCconductivity}.  

As just observed in the context of the DC electric conductivity, these more general gravitational duals open the path for a much richer phenomenology, specially in the thermal transport sector which is strongly modified by the Horndeski terms. In particular, one could think to modify not only the gravity/scalar sectors but also the Maxwell one and to consider more general asymptotics such as Lifshitz geometries, or geometries with hyperscaling violation. A complete analysis of the phenomenology of the dual field theories goes beyond the scope of this work and it will be considered in the near future.

\section{Conclusions}
\label{sec6}
Despite the successes of GR, there are still plenty of reasons to reasonably modify the cornerstone of gravitation, the Einstein-Hilbert action. From the cosmological constant problem and the questionable validity of GR in the very deep infrared (that is, at very large distances), up to quantum gravity corrections and bottom-up holographic applications, looking for robust alternatives is a justified course of action, and in this direction, scalar-tensor theories are perhaps the most prominent examples that do the job in an absolutely essential manner, i.e., by introducing an additional scalar degree of freedom which may or may not couple non-minimally to gravity. In particular, higher-derivative self-interactions prove to be quite interesting deformations, but they usually involve the risk of ending up with an unstable theory. This endogenous danger can certainly be avoided if the higher-derivative contributions somehow cancel out at the level of the field equations, which is exactly the idea behind the theory of Horndeski~\cite{Horndeski:1974wa} and its re-emergence in the context of Galileons~\cite{Deffayet:2009wt}; by adding appropriate gravitational counterterms, one may suppress the appearance of higher-derivative terms in the equations of motion, keeping them up to second-order in derivatives of the fields. 

Still, just putting forth an action principle, although important in its own right, is for various reasons not enough, and one generally wishes to associate a physically meaningful and interesting solution spectrum with it. However, the task of finding exact solutions to a variationally obtained set of field equations is usually a highly non-trivial one, with its complexity depending on the complexity of the model under study and the amount of symmetry being introduced. In this regard, Horndeski theory poses a significant challenge, especially if one is looking for exact black hole solutions. As expected, this road is paved with no-go theorems, dead ends, intractable expressions and naked singularities. Nonetheless, if one manages to overcome these obstacles, a rewarding plethora of highly interesting black hole solutions may be found. Yet, up to today, a large region of the ``black'' landscape of Horndeski gravity surely remains unexplored, mainly because of the technical difficulties involved, part of which also has to do with the fact that the scalar field is usually taken to depend on the radial coordinate $r$, a statement compatible with staticity and spherical symmetry. 

In the current work, we consider the Einstein-Maxwell theory deformed by the truncated version~\eqref{eq:HorndeskiAction} of shift- and reflection-symmetric Horndeski gravity. We show that for backgrounds of the form~\eqref{eq:ScalarBackgrounds}, the difficulty of having to integrate the scalar equations of motion is completely lifted, as the latter are identically satisfied. Note, however, that this happens at the cost of reducing the full symmetry of~\eqref{eq:TotalAction} down to a residual diagonal subgroup, thus, also partially breaking diffeomorphism invariance in the bulk. For this fairly general class of theories, and starting from a minimal set of requirements found in the beginning of section~\ref{sec:Algo}, we provide a systematic way of integrating the metric equations of motion which is schematically given in figure~\ref{fig:FlowchartFig}. This has the form of a \emph{solution-generating algorithm}, a step-sequence which we divide it into three separate phases. The algorithm is meant as a means of finding exact solutions respecting these requirements; in particular, this means that the integrals in~\eqref{eq:MasterEquationC},~\eqref{eq:BetterFormf} and~\eqref{eq:kEquation} must have a closed-form expression, otherwise the whole machinery will not have the desired result. The success of this method relies on an ``ad hoc'' change of variables, from $r$ to $\bar{X}$, with no appearance of the unknown metric functions in the inversion formula~\eqref{eq:rofX} because of the precise backgrounds~\eqref{eq:ScalarBackgrounds}. In terms of $\bar{X}$, the field equations become significantly more tractable, and a straightforward integration can be carried out in a convenient sequence of steps (see the detailed phases analysis in the end of section~\ref{sec:Algo}).

Some interesting ``universal'' results can be also reported at an early stage without specifying the form of $G_4$ and $K$. Assuming the solutions describe black hole configurations, we show that one can obtain expressions for the Hawking temperature and Wald entropy density also in terms of the Horndeski functions, and some first restrictions can be placed. Additionally, we argue that if zero-temperature configurations exist, their size will not be affected by $G_4$, although the latter contributes to their mass. We further solve the field equations near a hypothetical extremal horizon for an $AdS_2\times \mathbb{R}^2$ IR fixed point, and display the explicit dependence of the radius of $AdS_2$ on the Horndeski functions. All extremal bulk solutions we present end up interpolating between this IR geometry and some model-dependent UV counterpart. 

Due to the form of the algorithm, we argue that it makes sense to classify solutions based on $G_4$. A preliminary exploration of the black hole spectrum is performed by putting our recipe to use, where we immediately (and rapidly) disclose new families of planar black holes in example classes $G_4\propto \bar X$ and $G_4\propto (-\bar X)^m$. In both cases, before fixing the form of k-essence, we do a preliminary asymptotic analysis to filter out choices of $K$ that lead to certain unwanted scenarios. Then, taking into account these restrictions, we choose a $K$, and derive novel exact black branes with axionic hair. Note that the families we find contain the solutions in~\cite{Jiang:2017imk,Figueroa:2020tya}. Apart from AAdS black branes, we derive a particularly interesting Lifshitz-like family in class $G_4\propto (-\bar{X})^m$. To the best of our knowledge, this is the first family of exact asymptotically Lifshitz black branes in Horndeski theory with $z>1$, where $z$ is a function of $m$. 

Finally, in the context of applied bottom-up holography, we start a preliminary investigation of the transport properties of the dual field theory in the case of an AdS boundary. A detailed derivation of the transport matrix is presented for arbitrary Horndeski functions (also in the presence of a time-dependent heat source), where boundary conditions for the perturbations are discussed in detail. We show, using the example of the DC electric conductivity, that this class of models could give rise to a very rich phenomenology (especially in the strongly modified heat sector) which could be potentially relevant to tackle existing open questions in the condensed matter ballpark.

Needless to say, there are several interesting future directions that can be pursued. For example, an appealing course of action would be to reconstruct the algorithm in higher spacetime dimensions, where a general second-order scalar tensor theory would indeed possess a richer quartic sector, taking into account that, for example, higher terms in the Lanczos-Lovelock series could be also introduced for appropriate $d$. Additionally, either for scalar-tensor or vector-tensor theories, extensions with higher-order field equations can be also studied; this is the case of, e.g., DHOST theories \cite{Langlois:2015cwa,Langlois:2015skt}, in which, although the theory is higher-order, no Ostrogradsky mode propagates due to the degeneracy of the corresponding kinetic matrix. In the latter class, solutions with a non-constant kinetic term for the scalar field are hard to find \cite{Minamitsuji:2019tet,Babichev:2020qpr}, and in this direction, it would be interesting to see if exporting the setup/method in this work to such cases would bear fruit. If solutions are indeed to be found, then one expects that they will be related to the ones disclosed here by means of a disformal transformation, nevertheless with a completely different causal structure. Moreover, it would be worth exploring if solutions with a purely time-dependent Galileon background of the form $\phi \sim t$, reminiscent of the ghost-condensate background~\cite{ArkaniHamed:2003uy}, could be realized without occurring in evident pathological behaviours; perhaps more interesting in this scenario, would be the holographic interpretation of broken time translations and the possible phenomenological implications. Finally, it would be important to provide a complete characterization of the transport properties of the dual field theories and extract further universal features which are independent of a precise choice of the various Horndeski functions.  These questions go beyond the investigation performed in this work and are left to be studied in the future.

\section*{Acknowledgments}
A. C. would like to thank Jose Barrientos and Julio Oliva for useful discussions on related topics. M. B. acknowledges the support of the  Shanghai Municipal Science and Technology Major Project (Grant No.2019SHZDZX01). A. C. work is funded by FONDECYT grant 1210500. K. P. acknowledges financial support provided by the European Regional Development Fund (ERDF) through the Center of Excellence TK133 ``The Dark Side of the Universe'' and PRG356 ``Gauge gravity: unification, extensions and phenomenology''.
\appendix

\bibliographystyle{JHEP}
\bibliography{Refs}

\providecommand{\href}[2]{#2}\begingroup\raggedright\begin{thebibliography}{10}

\bibitem{Maldacena:1997re}
J.~M. Maldacena, {\it {The Large N limit of superconformal field theories and
  supergravity}},  {\em Adv. Theor. Math. Phys.} {\bf 2} (1998) 231--252,
  [\href{http://arxiv.org/abs/hep-th/9711200}{{\tt hep-th/9711200}}].

\bibitem{CasalderreySolana:2011us}
J.~Casalderrey-Solana, H.~Liu, D.~Mateos, K.~Rajagopal, and U.~A. Wiedemann,
  {\em {Gauge/String Duality, Hot QCD and Heavy Ion Collisions}}.
\newblock Cambridge University Press, 2014.

\bibitem{Hartnoll:2009sz}
S.~A. Hartnoll, {\it {Lectures on holographic methods for condensed matter
  physics}},  {\em Class. Quant. Grav.} {\bf 26} (2009) 224002,
  [\href{http://arxiv.org/abs/0903.3246}{{\tt arXiv:0903.3246}}].

\bibitem{Hartnoll:2016apf}
S.~A. Hartnoll, A.~Lucas, and S.~Sachdev, {\it {Holographic quantum matter}},
  \href{http://arxiv.org/abs/1612.07324}{{\tt arXiv:1612.07324}}.

\bibitem{Hartnoll:2008vx}
S.~A. Hartnoll, C.~P. Herzog, and G.~T. Horowitz, {\it {Building a Holographic
  Superconductor}},  {\em Phys. Rev. Lett.} {\bf 101} (2008) 031601,
  [\href{http://arxiv.org/abs/0803.3295}{{\tt arXiv:0803.3295}}].

\bibitem{Bardoux:2012aw}
Y.~Bardoux, M.~M. Caldarelli, and C.~Charmousis, {\it {Shaping black holes with
  free fields}},  {\em JHEP} {\bf 05} (2012) 054,
  [\href{http://arxiv.org/abs/1202.4458}{{\tt arXiv:1202.4458}}].

\bibitem{zaanen2015holographic}
J.~Zaanen, Y.~Liu, Y.~Sun, and K.~Schalm, {\em Holographic Duality in Condensed
  Matter Physics}.
\newblock Cambridge University Press, 2015.

\bibitem{Baggioli:2019rrs}
M.~Baggioli, {\em {Applied Holography}: {A Practical Mini-Course}}.
\newblock SpringerBriefs in Physics. Springer, 2019.

\bibitem{Vegh:2013sk}
D.~Vegh, {\it {Holography without translational symmetry}},
  \href{http://arxiv.org/abs/1301.0537}{{\tt arXiv:1301.0537}}.

\bibitem{Alberte:2015isw}
L.~Alberte, M.~Baggioli, A.~Khmelnitsky, and O.~Pujolas, {\it {Solid Holography
  and Massive Gravity}},  {\em JHEP} {\bf 02} (2016) 114,
  [\href{http://arxiv.org/abs/1510.09089}{{\tt arXiv:1510.09089}}].

\bibitem{Blake:2013owa}
M.~Blake, D.~Tong, and D.~Vegh, {\it {Holographic Lattices Give the Graviton an
  Effective Mass}},  {\em Phys. Rev. Lett.} {\bf 112} (2014), no.~7 071602,
  [\href{http://arxiv.org/abs/1310.3832}{{\tt arXiv:1310.3832}}].

\bibitem{Hinterbichler:2011tt}
K.~Hinterbichler, {\it {Theoretical Aspects of Massive Gravity}},  {\em Rev.
  Mod. Phys.} {\bf 84} (2012) 671--710,
  [\href{http://arxiv.org/abs/1105.3735}{{\tt arXiv:1105.3735}}].

\bibitem{Dubovsky:2004sg}
S.~L. Dubovsky, {\it {Phases of massive gravity}},  {\em JHEP} {\bf 10} (2004)
  076, [\href{http://arxiv.org/abs/hep-th/0409124}{{\tt hep-th/0409124}}].

\bibitem{Mateos:2011ix}
D.~Mateos and D.~Trancanelli, {\it {The anisotropic N=4 super Yang-Mills plasma
  and its instabilities}},  {\em Phys. Rev. Lett.} {\bf 107} (2011) 101601,
  [\href{http://arxiv.org/abs/1105.3472}{{\tt arXiv:1105.3472}}].

\bibitem{Ge:2014aza}
X.-H. Ge, Y.~Ling, C.~Niu, and S.-J. Sin, {\it {Thermoelectric conductivities,
  shear viscosity, and stability in an anisotropic linear axion model}},  {\em
  Phys. Rev. D} {\bf 92} (2015), no.~10 106005,
  [\href{http://arxiv.org/abs/1412.8346}{{\tt arXiv:1412.8346}}].

\bibitem{Andrade:2013gsa}
T.~Andrade and B.~Withers, {\it {A simple holographic model of momentum
  relaxation}},  {\em JHEP} {\bf 05} (2014) 101,
  [\href{http://arxiv.org/abs/1311.5157}{{\tt arXiv:1311.5157}}].

\bibitem{Baggioli:2014roa}
M.~Baggioli and O.~Pujolas, {\it {Electron-Phonon Interactions, Metal-Insulator
  Transitions, and Holographic Massive Gravity}},  {\em Phys. Rev. Lett.} {\bf
  114} (2015), no.~25 251602, [\href{http://arxiv.org/abs/1411.1003}{{\tt
  arXiv:1411.1003}}].

\bibitem{Baggioli:2021xuv}
M.~Baggioli, K.-Y. Kim, L.~Li, and W.-J. Li, {\it {Holographic Axion Model: a
  simple gravitational tool for quantum matter}},
  \href{http://arxiv.org/abs/2101.01892}{{\tt arXiv:2101.01892}}.

\bibitem{Alberte:2017oqx}
L.~Alberte, M.~Ammon, A.~Jim\'enez-Alba, M.~Baggioli, and O.~Pujol\`as, {\it
  {Holographic Phonons}},  {\em Phys. Rev. Lett.} {\bf 120} (2018), no.~17
  171602, [\href{http://arxiv.org/abs/1711.03100}{{\tt arXiv:1711.03100}}].

\bibitem{Alberte:2017cch}
L.~Alberte, M.~Ammon, M.~Baggioli, A.~Jim\'enez, and O.~Pujol\`as, {\it {Black
  hole elasticity and gapped transverse phonons in holography}},  {\em JHEP}
  {\bf 01} (2018) 129, [\href{http://arxiv.org/abs/1708.08477}{{\tt
  arXiv:1708.08477}}].

\bibitem{Ammon:2019wci}
M.~Ammon, M.~Baggioli, and A.~Jim\'enez-Alba, {\it {A Unified Description of
  Translational Symmetry Breaking in Holography}},  {\em JHEP} {\bf 09} (2019)
  124, [\href{http://arxiv.org/abs/1904.05785}{{\tt arXiv:1904.05785}}].

\bibitem{Davison:2013txa}
R.~A. Davison, K.~Schalm, and J.~Zaanen, {\it {Holographic duality and the
  resistivity of strange metals}},  {\em Phys. Rev. B} {\bf 89} (2014), no.~24
  245116, [\href{http://arxiv.org/abs/1311.2451}{{\tt arXiv:1311.2451}}].

\bibitem{Blake:2014yla}
M.~Blake and A.~Donos, {\it {Quantum Critical Transport and the Hall Angle}},
  {\em Phys. Rev. Lett.} {\bf 114} (2015), no.~2 021601,
  [\href{http://arxiv.org/abs/1406.1659}{{\tt arXiv:1406.1659}}].

\bibitem{Amoretti:2016cad}
A.~Amoretti, M.~Baggioli, N.~Magnoli, and D.~Musso, {\it {Chasing the cuprates
  with dilatonic dyons}},  {\em JHEP} {\bf 06} (2016) 113,
  [\href{http://arxiv.org/abs/1603.03029}{{\tt arXiv:1603.03029}}].

\bibitem{Hoyos:2010at}
C.~Hoyos and P.~Koroteev, {\it {On the Null Energy Condition and Causality in
  Lifshitz Holography}},  {\em Phys. Rev. D} {\bf 82} (2010) 084002,
  [\href{http://arxiv.org/abs/1007.1428}{{\tt arXiv:1007.1428}}]. [Erratum:
  Phys.Rev.D 82, 109905 (2010)].

\bibitem{PhysRevB.91.155126}
S.~A. Hartnoll and A.~Karch, {\it Scaling theory of the cuprate strange
  metals},  {\em Phys. Rev. B} {\bf 91} (Apr, 2015) 155126.

\bibitem{Taylor:2015glc}
M.~Taylor, {\it {Lifshitz holography}},  {\em Class. Quant. Grav.} {\bf 33}
  (2016), no.~3 033001, [\href{http://arxiv.org/abs/1512.03554}{{\tt
  arXiv:1512.03554}}].

\bibitem{Chemissany:2014xsa}
W.~Chemissany and I.~Papadimitriou, {\it {Lifshitz holography: The whole
  shebang}},  {\em JHEP} {\bf 01} (2015) 052,
  [\href{http://arxiv.org/abs/1408.0795}{{\tt arXiv:1408.0795}}].

\bibitem{Horndeski:1974wa}
G.~W. Horndeski, {\it {Second-order scalar-tensor field equations in a
  four-dimensional space}},  {\em Int. J. Theor. Phys.} {\bf 10} (1974)
  363--384.

\bibitem{Baggioli:2017ojd}
M.~Baggioli and W.-J. Li, {\it {Diffusivities bounds and chaos in holographic
  Horndeski theories}},  {\em JHEP} {\bf 07} (2017) 055,
  [\href{http://arxiv.org/abs/1705.01766}{{\tt arXiv:1705.01766}}].

\bibitem{Figueroa:2020tya}
J.~P. Figueroa and K.~Pallikaris, {\it {Quartic Horndeski, planar black holes,
  holographic aspects and universal bounds}},  {\em JHEP} {\bf 20} (2020) 090,
  [\href{http://arxiv.org/abs/2006.00967}{{\tt arXiv:2006.00967}}].

\bibitem{Cisterna:2017jmv}
A.~Cisterna, M.~Hassaine, J.~Oliva, and M.~Rinaldi, {\it {Axionic black branes
  in the k-essence sector of the Horndeski model}},  {\em Phys. Rev. D} {\bf
  96} (2017), no.~12 124033, [\href{http://arxiv.org/abs/1708.07194}{{\tt
  arXiv:1708.07194}}].

\bibitem{Cisterna:2018hzf}
A.~Cisterna, C.~Erices, X.-M. Kuang, and M.~Rinaldi, {\it {Axionic black branes
  with conformal coupling}},  {\em Phys. Rev. D} {\bf 97} (2018), no.~12
  124052, [\href{http://arxiv.org/abs/1803.07600}{{\tt arXiv:1803.07600}}].

\bibitem{Cisterna:2019uek}
A.~Cisterna, L.~Guajardo, and M.~Hassaine, {\it {Axionic charged black branes
  with arbitrary scalar nonminimal coupling}},  {\em Eur. Phys. J. C} {\bf 79}
  (2019), no.~5 418, [\href{http://arxiv.org/abs/1901.00514}{{\tt
  arXiv:1901.00514}}]. [Erratum: Eur.Phys.J.C 79, 710 (2019)].

\bibitem{Liu:2017kml}
H.-S. Liu, H.~Lu, and C.~N. Pope, {\it {Holographic Heat Current as Noether
  Current}},  {\em JHEP} {\bf 09} (2017) 146,
  [\href{http://arxiv.org/abs/1708.02329}{{\tt arXiv:1708.02329}}].

\bibitem{Jiang:2017imk}
W.-J. Jiang, H.-S. Liu, H.~Lu, and C.~N. Pope, {\it {DC Conductivities with
  Momentum Dissipation in Horndeski Theories}},  {\em JHEP} {\bf 07} (2017)
  084, [\href{http://arxiv.org/abs/1703.00922}{{\tt arXiv:1703.00922}}].

\bibitem{Liu:2016njg}
H.-S. Liu, H.~Lu, and C.~N. Pope, {\it {Magnetically-Charged Black Branes and
  Viscosity/Entropy Ratios}},  {\em JHEP} {\bf 12} (2016) 097,
  [\href{http://arxiv.org/abs/1602.07712}{{\tt arXiv:1602.07712}}].

\bibitem{Feng:2015wvb}
X.-H. Feng, H.-S. Liu, H.~L\"u, and C.~N. Pope, {\it {Thermodynamics of Charged
  Black Holes in Einstein-Horndeski-Maxwell Theory}},  {\em Phys. Rev. D} {\bf
  93} (2016), no.~4 044030, [\href{http://arxiv.org/abs/1512.02659}{{\tt
  arXiv:1512.02659}}].

\bibitem{Feng:2015oea}
X.-H. Feng, H.-S. Liu, H.~L\"u, and C.~N. Pope, {\it {Black Hole Entropy and
  Viscosity Bound in Horndeski Gravity}},  {\em JHEP} {\bf 11} (2015) 176,
  [\href{http://arxiv.org/abs/1509.07142}{{\tt arXiv:1509.07142}}].

\bibitem{Deffayet:2011gz}
C.~Deffayet, X.~Gao, D.~A. Steer, and G.~Zahariade, {\it {From k-essence to
  generalised Galileons}},  {\em Phys. Rev. D} {\bf 84} (2011) 064039,
  [\href{http://arxiv.org/abs/1103.3260}{{\tt arXiv:1103.3260}}].

\bibitem{Deffayet:2013lga}
C.~Deffayet and D.~A. Steer, {\it {A formal introduction to Horndeski and
  Galileon theories and their generalizations}},  {\em Class. Quant. Grav.}
  {\bf 30} (2013) 214006, [\href{http://arxiv.org/abs/1307.2450}{{\tt
  arXiv:1307.2450}}].

\bibitem{Kobayashi:2011nu}
T.~Kobayashi, M.~Yamaguchi, and J.~Yokoyama, {\it {Generalized G-inflation:
  Inflation with the most general second-order field equations}},  {\em Prog.
  Theor. Phys.} {\bf 126} (2011) 511--529,
  [\href{http://arxiv.org/abs/1105.5723}{{\tt arXiv:1105.5723}}].

\bibitem{Nicolis:2015sra}
A.~Nicolis, R.~Penco, F.~Piazza, and R.~Rattazzi, {\it {Zoology of condensed
  matter: Framids, ordinary stuff, extra-ordinary stuff}},  {\em JHEP} {\bf 06}
  (2015) 155, [\href{http://arxiv.org/abs/1501.03845}{{\tt arXiv:1501.03845}}].

\bibitem{Nicolis:2013lma}
A.~Nicolis, R.~Penco, and R.~A. Rosen, {\it {Relativistic Fluids, Superfluids,
  Solids and Supersolids from a Coset Construction}},  {\em Phys. Rev. D} {\bf
  89} (2014), no.~4 045002, [\href{http://arxiv.org/abs/1307.0517}{{\tt
  arXiv:1307.0517}}].

\bibitem{Alberte:2018doe}
L.~Alberte, M.~Baggioli, V.~C. Castillo, and O.~Pujolas, {\it {Elasticity
  bounds from Effective Field Theory}},  {\em Phys. Rev. D} {\bf 100} (2019),
  no.~6 065015, [\href{http://arxiv.org/abs/1807.07474}{{\tt
  arXiv:1807.07474}}]. [Erratum: Phys.Rev.D 102, 069901 (2020)].

\bibitem{Caldarelli:2016nni}
M.~M. Caldarelli, A.~Christodoulou, I.~Papadimitriou, and K.~Skenderis, {\it
  {Phases of planar AdS black holes with axionic charge}},  {\em JHEP} {\bf 04}
  (2017) 001, [\href{http://arxiv.org/abs/1612.07214}{{\tt arXiv:1612.07214}}].

\bibitem{Hervik:2019gly}
S.~Hervik and M.~Ortaggio, {\it {Universal Black Holes}},  {\em JHEP} {\bf 02}
  (2020) 047, [\href{http://arxiv.org/abs/1907.08788}{{\tt arXiv:1907.08788}}].

\bibitem{Hervik:2020nxs}
S.~Hervik and M.~Ortaggio, {\it {On Universal Black Holes}},  {\em Acta Phys.
  Polon. Supp.} {\bf 13} (2020) 291,
  [\href{http://arxiv.org/abs/2007.00556}{{\tt arXiv:2007.00556}}].

\bibitem{Hervik:2020zvn}
S.~Hervik and M.~Ortaggio, {\it {Universal $p$-form black holes in generalized
  theories of gravity}},  {\em Eur. Phys. J. C} {\bf 80} (2020), no.~11 1020,
  [\href{http://arxiv.org/abs/2007.05464}{{\tt arXiv:2007.05464}}].

\bibitem{Wald:1993nt}
R.~M. Wald, {\it {Black hole entropy is the Noether charge}},  {\em Phys. Rev.
  D} {\bf 48} (1993), no.~8 R3427--R3431,
  [\href{http://arxiv.org/abs/gr-qc/9307038}{{\tt gr-qc/9307038}}].

\bibitem{Brustein:2007jj}
R.~Brustein, D.~Gorbonos, and M.~Hadad, {\it {Wald's entropy is equal to a
  quarter of the horizon area in units of the effective gravitational
  coupling}},  {\em Phys. Rev. D} {\bf 79} (2009) 044025,
  [\href{http://arxiv.org/abs/0712.3206}{{\tt arXiv:0712.3206}}].

\bibitem{abramowitz+stegun}
M.~Abramowitz and I.~A. Stegun, {\em Handbook of Mathematical Functions with
  Formulas, Graphs, and Mathematical Tables}.
\newblock Dover, New York, ninth dover printing, tenth gpo printing~ed., 1964.

\bibitem{Copsey:2012gw}
K.~Copsey and R.~Mann, {\it {Singularities in Hyperscaling Violating
  Spacetimes}},  {\em JHEP} {\bf 04} (2013) 079,
  [\href{http://arxiv.org/abs/1210.1231}{{\tt arXiv:1210.1231}}].

\bibitem{Martinez:2005di}
C.~Martinez, J.~P. Staforelli, and R.~Troncoso, {\it {Topological black holes
  dressed with a conformally coupled scalar field and electric charge}},  {\em
  Phys. Rev. D} {\bf 74} (2006) 044028,
  [\href{http://arxiv.org/abs/hep-th/0512022}{{\tt hep-th/0512022}}].

\bibitem{Hartnoll:2020fhc}
S.~A. Hartnoll, G.~T. Horowitz, J.~Kruthoff, and J.~E. Santos, {\it {Diving
  into a holographic superconductor}},  {\em SciPost Phys.} {\bf 10} (2021)
  009, [\href{http://arxiv.org/abs/2008.12786}{{\tt arXiv:2008.12786}}].

\bibitem{Cai:2020wrp}
R.-G. Cai, L.~Li, and R.-Q. Yang, {\it {No Inner-Horizon Theorem for Black
  Holes with Charged Scalar Hairs}},  {\em JHEP} {\bf 03} (2021) 263,
  [\href{http://arxiv.org/abs/2009.05520}{{\tt arXiv:2009.05520}}].

\bibitem{Grandi:2021ajl}
N.~Grandi and I.~Salazar~Landea, {\it {Diving inside a hairy black hole}},
  \href{http://arxiv.org/abs/2102.02707}{{\tt arXiv:2102.02707}}.

\bibitem{Yang:2021civ}
R.-Q. Yang, R.-G. Cai, and L.~Li, {\it {Constraining the number of horizons
  with energy conditions}},  \href{http://arxiv.org/abs/2104.03012}{{\tt
  arXiv:2104.03012}}.

\bibitem{Devecioglu:2021xug}
D.~O. Devecioglu and M.-I. Park, {\it {No Scalar-Haired Cauchy Horizon Theorem
  in Einstein-Maxwell-Horndeski Theories}},
  \href{http://arxiv.org/abs/2101.10116}{{\tt arXiv:2101.10116}}.

\bibitem{Kachru:2008yh}
S.~Kachru, X.~Liu, and M.~Mulligan, {\it {Gravity duals of Lifshitz-like fixed
  points}},  {\em Phys. Rev. D} {\bf 78} (2008) 106005,
  [\href{http://arxiv.org/abs/0808.1725}{{\tt arXiv:0808.1725}}].

\bibitem{2013arXiv1312.7736B}
M.~{Bravo-Gaete} and M.~{Hassaine}, {\it {Lifshitz black holes with a
  time-dependent scalar field in Horndeski theory}},  {\em arXiv e-prints}
  (Dec., 2013) arXiv:1312.7736, [\href{http://arxiv.org/abs/1312.7736}{{\tt
  arXiv:1312.7736}}].

\bibitem{Brito:2019ose}
F.~A. Brito and F.~F. Santos, {\it {Black brane in asymptotically Lifshitz
  spacetime and viscosity/entropy ratios in Horndeski gravity}},  {\em EPL}
  {\bf 129} (2020), no.~5 50003, [\href{http://arxiv.org/abs/1901.06770}{{\tt
  arXiv:1901.06770}}].

\bibitem{Donos:2014cya}
A.~Donos and J.~P. Gauntlett, {\it {Thermoelectric DC conductivities from black
  hole horizons}},  {\em JHEP} {\bf 11} (2014) 081,
  [\href{http://arxiv.org/abs/1406.4742}{{\tt arXiv:1406.4742}}].

\bibitem{Davison:2013jba}
R.~A. Davison, {\it {Momentum relaxation in holographic massive gravity}},
  {\em Phys. Rev. D} {\bf 88} (2013) 086003,
  [\href{http://arxiv.org/abs/1306.5792}{{\tt arXiv:1306.5792}}].

\bibitem{Grozdanov:2015qia}
S.~Grozdanov, A.~Lucas, S.~Sachdev, and K.~Schalm, {\it {Absence of
  disorder-driven metal-insulator transitions in simple holographic models}},
  {\em Phys. Rev. Lett.} {\bf 115} (2015), no.~22 221601,
  [\href{http://arxiv.org/abs/1507.00003}{{\tt arXiv:1507.00003}}].

\bibitem{Baggioli:2016oqk}
M.~Baggioli and O.~Pujolas, {\it {On holographic disorder-driven
  metal-insulator transitions}},  {\em JHEP} {\bf 01} (2017) 040,
  [\href{http://arxiv.org/abs/1601.07897}{{\tt arXiv:1601.07897}}].

\bibitem{Baggioli:2016oju}
M.~Baggioli and O.~Pujolas, {\it {On Effective Holographic Mott Insulators}},
  {\em JHEP} {\bf 12} (2016) 107, [\href{http://arxiv.org/abs/1604.08915}{{\tt
  arXiv:1604.08915}}].

\bibitem{Deffayet:2009wt}
C.~Deffayet, G.~Esposito-Farese, and A.~Vikman, {\it {Covariant Galileon}},
  {\em Phys. Rev. D} {\bf 79} (2009) 084003,
  [\href{http://arxiv.org/abs/0901.1314}{{\tt arXiv:0901.1314}}].

\bibitem{Langlois:2015cwa}
D.~Langlois and K.~Noui, {\it {Degenerate higher derivative theories beyond
  Horndeski: evading the Ostrogradski instability}},  {\em JCAP} {\bf 02}
  (2016) 034, [\href{http://arxiv.org/abs/1510.06930}{{\tt arXiv:1510.06930}}].

\bibitem{Langlois:2015skt}
D.~Langlois and K.~Noui, {\it {Hamiltonian analysis of higher derivative
  scalar-tensor theories}},  {\em JCAP} {\bf 07} (2016) 016,
  [\href{http://arxiv.org/abs/1512.06820}{{\tt arXiv:1512.06820}}].

\bibitem{Minamitsuji:2019tet}
M.~Minamitsuji and J.~Edholm, {\it {Black holes with a nonconstant kinetic term
  in degenerate higher-order scalar tensor theories}},  {\em Phys. Rev. D} {\bf
  101} (2020), no.~4 044034, [\href{http://arxiv.org/abs/1912.01744}{{\tt
  arXiv:1912.01744}}].

\bibitem{Babichev:2020qpr}
E.~Babichev, C.~Charmousis, A.~Cisterna, and M.~Hassaine, {\it {Regular black
  holes via the Kerr-Schild construction in DHOST theories}},  {\em JCAP} {\bf
  06} (2020) 049, [\href{http://arxiv.org/abs/2004.00597}{{\tt
  arXiv:2004.00597}}].

\bibitem{ArkaniHamed:2003uy}
N.~Arkani-Hamed, H.-C. Cheng, M.~A. Luty, and S.~Mukohyama, {\it {Ghost
  condensation and a consistent infrared modification of gravity}},  {\em JHEP}
  {\bf 05} (2004) 074, [\href{http://arxiv.org/abs/hep-th/0312099}{{\tt
  hep-th/0312099}}].

\end{thebibliography}\endgroup
\end{document}